\documentclass[sigconf,nonacm,screen]{acmart}

\usepackage{booktabs,colortbl,multicol,multirow}
\usepackage{fontawesome}
\usepackage[table]{xcolor}
\usepackage{pgf}

\newcommand{\gradcell}[1]{%
  \begingroup
  \pgfmathsetmacro{\val}{#1}%
  \def\cellshade{}%
  \ifdim\val pt>0pt
    \pgfmathtruncatemacro{\shade}{min(80,round(80*\val/0.75))}%
    \xdef\cellshade{\noexpand\cellcolor{poscolor!\shade!white}}%
  \else
    \ifdim\val pt<0pt

    \pgfmathtruncatemacro{\shade}{min(80,round(80*abs(\val)/0.75))}%
      \xdef\cellshade{\noexpand\cellcolor{negcolor!\shade!white}}%
    \fi
  \fi
  \endgroup
  \cellshade #1%
}

\definecolor{negcolor}{HTML}{CD3700}
\definecolor{poscolor}{HTML}{018571}

\definecolor{DarkBlue}{HTML}{00008B}
\definecolor{mscolor}{HTML}{01665e}
\definecolor{nmscolor}{HTML}{bf812d}
\definecolor{lgreen}{HTML}{ccece6}
\definecolor{dolive}{HTML}{308014}
\definecolor{purple}{HTML}{ae017e}
\definecolor{brickred}{HTML}{f03b20}

\usepackage[most]{tcolorbox}

\newtcolorbox{takeawaybox}{
    enhanced,
    colback=black!4,
    frame hidden,
    boxrule=0pt,
    borderline west={3pt}{0pt}{blue!75!black},
    left=3mm,
    right=3mm,
    top=2mm,
    bottom=2mm,
    sharp corners
}

\newif{\ifhidecomments}
  \hidecommentsfalse 
\ifhidecomments
    \newcommand{\agam}[1]{}
    \newcommand{\claire}[1]{}
    \newcommand{\eshwar}[1]{}
\else
    \newcommand{\agam}[1]{\textbf{\small\sffamily{\textcolor{teal}{[#1 -- Agam]}}}}
    \newcommand{\claire}[1]{\textbf{\small\sffamily{\textcolor{DarkBlue}{[#1 -- Claire]}}}}
    \newcommand{\eshwar}[1]{\textbf{\small\sffamily{\textcolor{brickred}{[#1 -- Eshwar]}}}}
  \fi

\newcommand{\para}[1]{\noindent\textbf{\textit{#1}~}}
\usepackage{subcaption}
\AtBeginDocument{%
  }

\setcopyright{acmlicensed}
\copyrightyear{2026}
\acmYear{2026}
\acmDOI{XXXXXXX.XXXXXXX}
\acmConference[Conference acronym 'XX]{Make sure to enter the correct
  conference title from your rights confirmation email}{2026}{Woodstock, NY}

\begin{document}
\sloppy
\title[The Wisdom of the Loudest: A Large-Scale Audit of Generative Search on Reddit]{The Wisdom of the Loudest: \\A Large-Scale Audit of Generative Search on Reddit}


\author{Agam Goyal}
\orcid{0009-0009-5989-2887}
\affiliation{%
  \institution{University of Illinois Urbana-Champaign}
  \city{Urbana}
  \state{Illinois}
  \country{USA}}
\email{agamg2@illinois.edu}

\author{Wang Claire}
\orcid{0009-0003-3562-055X}
\affiliation{%
  \institution{University of Illinois Urbana-Champaign}
  \city{Urbana}
  \state{IL}
  \country{USA}
}
\email{claire46@illinois.edu}

\author{Eshwar Chandrasekharan}
\orcid{0000-0002-7473-1418}
\affiliation{%
  \institution{University of Illinois Urbana-Champaign}
  \city{Urbana}
  \state{Illinois}
  \country{USA}}
\email{eshwar@illinois.edu}

\renewcommand{\shortauthors}{Goyal et al.}

\begin{abstract}
Online communities are valued not only for answers, but for the diversity of experiences and perspectives they contain. Generative search increasingly mediates access to this discourse, yet little is known about which community voices survive retrieval and synthesis. We audit Reddit Answers using 10{,}000 queries from 20 advice- and support-seeking communities, repeated three times to produce 30{,}000 answers over 14.68M comments. We find that differences across runs are driven primarily by retrieval, answers routinely combine evidence across communities, and selection strongly favors already-visible, top-level comments. Formal and directive language is more likely to be surfaced, while experiential voice is less likely to survive selection and is further weakened during synthesis, with first-person singular language declining sharply. These findings show that community-grounded generative search is not neutral summarization, and should be designed not only for relevance and fluency, but also for provenance, plurality, and legibility.
\end{abstract}



\begin{teaserfigure}
    \includegraphics[width=\textwidth]{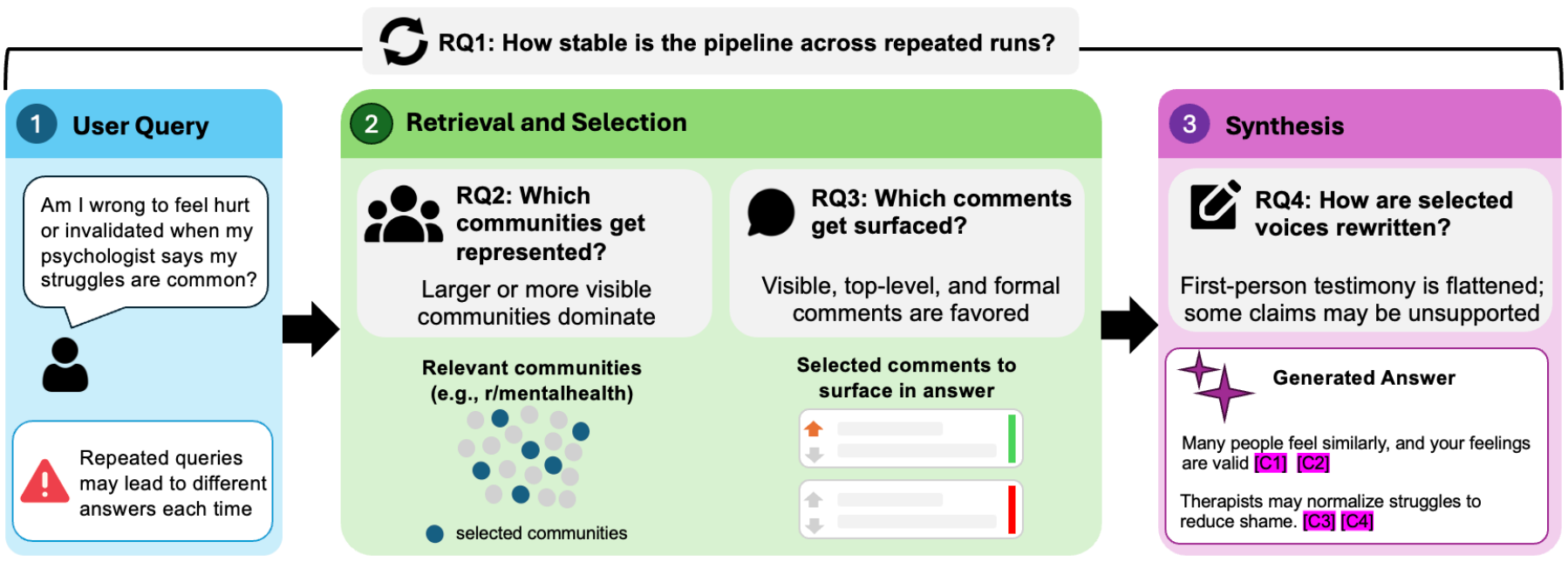}
    \caption{\textbf{Overview of this paper:} A user queries Reddit Answers, an AI-mediated search feature, which retrieves and selects evidence from across the platform before synthesizing it into a single response. We audit two complementary aspects of this process: which communities are represented in the retrieved evidence (\textbf{RQ2}), and which comments are ultimately surfaced in the answer (\textbf{RQ3}). We then examine how the selected evidence is rewritten during synthesis (\textbf{RQ4}). Finally, by repeating identical queries, we measure the stability of this entire process and identify where run-to-run variation arises (\textbf{RQ1}).}
    \Description{Overview of the Reddit Answers audit pipeline. A user query enters the system, followed by a combined retrieval and selection stage and then synthesis into a generated answer. The figure highlights four research questions. RQ1 spans the full pipeline and asks how stable the system is across repeated runs. RQ2 examines which communities are represented in retrieved evidence, with larger or more visible communities potentially favored. RQ3 examines which comments are surfaced, highlighting preferences for visible, top-level, and formal comments. RQ4 examines how selected voices are rewritten during synthesis, including the loss of first-person testimony and cases where generated claims may not be supported by cited evidence. The figure does not assume a fixed ordering between community and comment selection.}
    \label{fig:reddit-answers-pipeline}
\end{teaserfigure}


\maketitle

\section{Introduction}

\begin{center}
\begin{minipage}{0.78\linewidth}
    \vspace{6pt}
    \large\itshape
    ``Isn't the point of this place that there are real people to talk to??''

    \normalfont\normalsize
    \raggedleft
    ---Reddit user on the launch of Reddit Answers, \\Dec 2024 \textit{(paraphrased)}
\end{minipage}
\end{center}

\medskip
When Reddit introduced Reddit Answers, an AI-mediated platform-wide search, it described the system as a faster way to find ``human perspectives'' from conversations across the platform.\footnote{\href{https://redditinc.com/news/introducing-reddit-answers}{https://redditinc.com/news/introducing-reddit-answers}} Reddit users immediately questioned the premise, with one prominent reply asking what the tool offered over simply adding ``Reddit'' to a web search, while many dismissed it as ``AI garbage'' or ``worthless AI slop.''\footnote{\href{https://www.reddit.com/r/reddit/comments/1habm06/for_all_your_questions_introducing_reddit_answers/}{https://www.reddit.com/r/reddit/comments/1habm06/for\_all\_your\_questions\\\_introducing\_reddit\_answers/}} These reactions raise a deeper question than whether another AI search interface is useful. Once a system turns conversations among many people into a single answer, \textit{which communities and voices does it carry forward, which does it leave behind, and how does it transform those that survive?}

This question matters because every day, millions of people turn to online communities for things they cannot easily get from official sources or from people they already know: what a diagnosis feels like week-to-week, whether a financial decision worked out for someone in similar circumstances, or how another parent or partner handled a situation like their own~\cite{constant1996kindness,granovetter1977strength,ellison2007benefits,de2014seeking}. Platforms like Reddit host these exchanges across thousands of communities~\cite{gottfried2025americans}. Much of this activity involves advice and support seeking, where a question receives layered responses containing competing recommendations, evidence, stances, and first-person accounts~\cite{de2014mental,andalibi2016understanding,andalibi2017sensitive,nambisan2011information}. The value of these communities therefore lies not only in reaching an answer, but in seeing a diversity of positions and reading experiences in the voice of the people who lived them~\cite{hong2004groups,sourati2026homogenizing,de2014mental,andalibi2016understanding}.

\begin{figure*}
    \centering
    \includegraphics[width=0.95\linewidth]{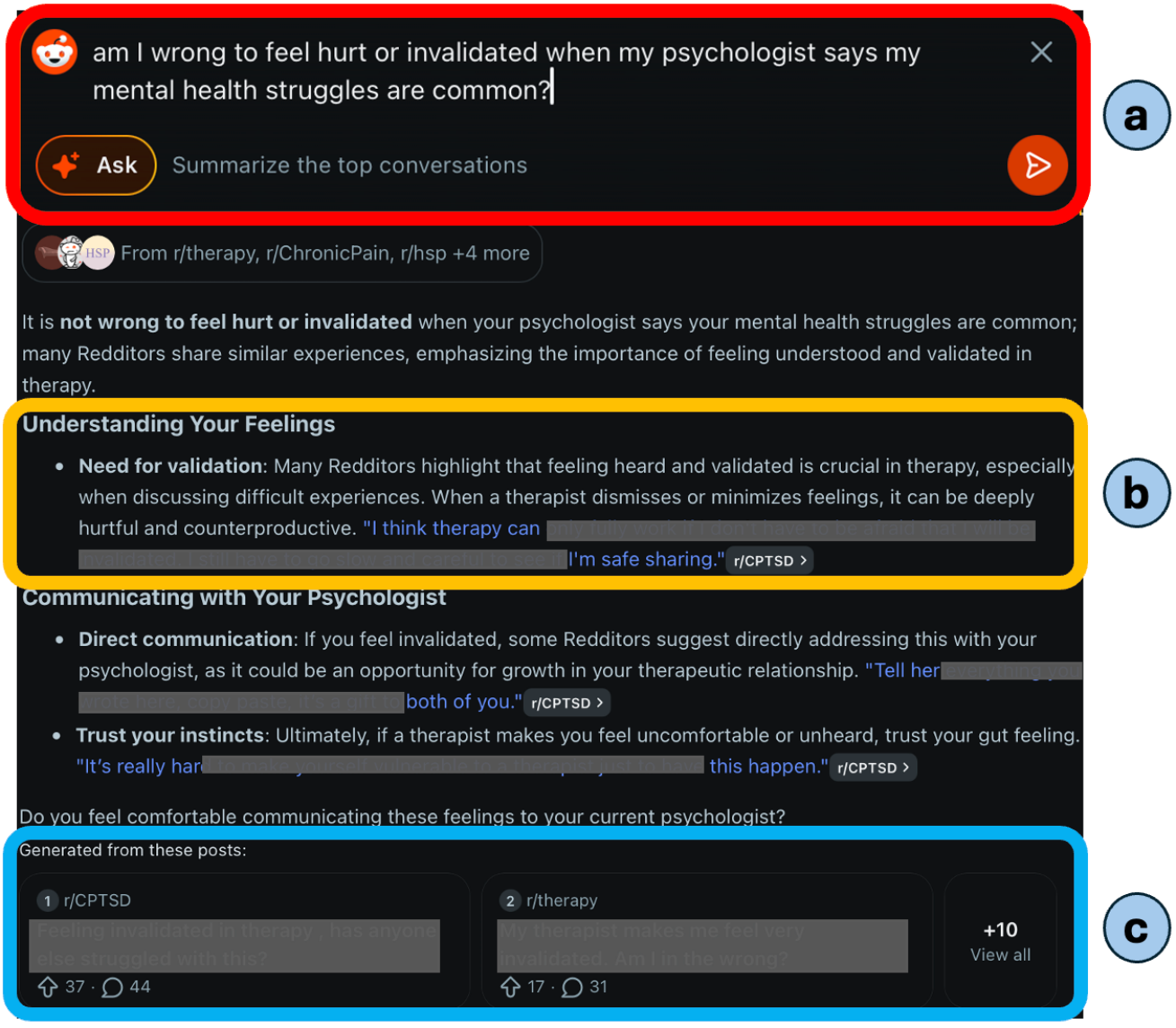}
    \caption{\textbf{Representational Anatomy of a Reddit Answers response:} (a) The user submits a free-text query with the \textit{``Ask''} AI-search mode enabled; (b) The system returns a sectioned synthesis that paraphrases retrieved community discussion and quotes comments inline. Each quotation carries a subreddit chip linking to its source comment; the surrounding prose speaks to the user in a narrator's voice; (c) A \textit{``Generated from these posts''} panel lists the retrieved source threads with their scores and comment counts.}
    \Description{Screenshot of the Reddit Answers interface divided into three highlighted regions. Region A at the top contains the user's free-text query and the Ask search mode. Region B contains the generated answer, organized into titled sections with explanatory prose and short quotations attributed to source comments through subreddit citation chips. Region C at the bottom is a ``Generated from these posts'' panel listing the Reddit threads used as sources, including post-level engagement information. The figure illustrates that the generated narrator text, quoted comments, and underlying source threads are separately visible in the interface.}
    \label{fig:reddit-answers-anatomy}
\end{figure*}

Today, \textit{generative search}~\cite{li2024survey,aral2026rise} increasingly sits between users and that discourse. Reddit Answers, an AI search system developed by Reddit, takes a user's query, retrieves posts and comments across the platform, and synthesizes them into a single sectioned response with inline quotations and links to its sources (see \autoref{fig:reddit-answers-anatomy}). This mode of interaction is fundamentally different from reading a thread. Instead of forming a judgment across multiple voices, the user receives one synthesized voice speaking on behalf of people across communities. The plurality of the underlying discussion can therefore be compressed into what appears to be a coherent consensus~\cite{padmakumar2024does,peterson2025ai}. 

Unlike web-grounded generative search systems whose retrieval pool spans the open web~\cite{perplexitySonar,searchGoogleAIO,bingIntroducingBing}, Reddit Answers draws specifically on community discourse. Its evidence comes from people participating in communities with distinct norms and accumulated expertise, and that discourse has already been shaped by voting, ranking, and moderation. Source selection therefore carries a different representational consequence. \textit{When the corpus is the platform's own discourse, whose contributions does AI-mediated synthesis amplify, and whose does it erase?}

A contribution's chance of being surfaced may depend on several properties of the underlying discussion. The system might favor community-conferred visibility, such as score or position in the thread. It might favor particular forms of writing, such as longer, more formal, or more prescriptive comments. Or it might favor authors with greater standing on the platform. Selection and synthesis effects of this kind have been documented in audits of other generative search systems~\cite{liu2023evaluating,venkit2024search,xu2026measuring,hu2026auditing,huang2026answer}, but community discourse makes their consequences particularly salient. Favoring visibility can inherit well-documented effects of early voting and ranking~\cite{muchnik2013social,salganik2006experimental,chan2025ranking}. Favoring particular linguistic registers can determine whether personal experiences are surfaced at all, and whether they remain recognizable as personal experiences when presented to users~\cite{huang2026answer,sourati2026homogenizing}. Favoring author standing can turn accumulated platform reputation into further exposure~\cite{josang2007survey}. Which signals survive these stages determines what kind of community knowledge ultimately reaches the user.

In this work, we conduct a large-scale audit of Reddit Answers using ecologically grounded queries distilled from the questions people actually bring to online communities. Starting from twelve months of posts across 20 advice- and support-seeking communities spanning four domains---finance, peer support, critical advice, and life decisions---we construct 10{,}000 queries, each linked back to the real community post from which it was distilled. We submit every query to Reddit Answers three times, yielding 30{,}000 generated responses. For the posts retrieved by the system, we collect their complete discussion threads through Reddit's official API, producing a corpus of 337{,}690 threads and 14.68M comments across 8{,}590 communities. This design lets us trace the system from repeated retrieval, through community routing and comment selection, to the final synthesized answer. We organize the audit around four research questions (see \autoref{fig:reddit-answers-pipeline}):

\noindent\textbf{RQ1 (Variation):} \textit{How much does a Reddit Answer vary from one run to the next?} We submit the same frozen query set three times back-to-back and decompose the variation we observe into the stage of the pipeline that produces it, which establishes what a single scrape can and cannot support.

\noindent\textbf{RQ2 (Routing):} \textit{Which communities do answers draw from?} We examine which communities the system draws on to answer a question, how often it draws on the community the question itself came from, and what explains the large differences between communities.
 
\noindent\textbf{RQ3 (Selection):} \textit{Which comments does Reddit Answers select and quote from the threads it retrieves?} We model selection within threads, asking how community-conferred visibility, comment form and register, and author-level platform capital shape whose contribution becomes part of the synthesized answer.

\noindent\textbf{RQ4 (Presentation):} \textit{How does synthesis transform the community's voice?} We compare the linguistic character of synthesized answers against the source comments they draw on, and against the answers a general web-grounded search system gives to the same queries. 

\subsubsection*{Summary of findings:}
We find that Reddit Answers does not simply recover and compress community discourse. Retrieval is unstable across repeated queries and accounts for most run-to-run variation in the final answer. The system routinely assembles evidence across community boundaries, with routing shaped by community size, prior visibility, and how distinctive a post is within its local context. Within retrieved threads, selection strongly favors comments that are already prominent, while formal and directive language is more likely to survive than experiential and supportive language. Synthesis then further weakens first-person cues, making community testimony less recognizable as testimony by the time it reaches the user.

\subsubsection*{Summary of implications:}
These findings suggest that community-grounded generative search should be evaluated as a representational pipeline rather than only as a question-answering system. Improving the final generator is insufficient if upstream retrieval and selection have already narrowed which communities, perspectives, and forms of evidence are available to summarize. Systems built over community discourse should therefore make provenance, disagreement, and source context more legible, and should consider retrieval objectives that preserve substantively different perspectives rather than only the most visible or answer-like contributions. More broadly, audits of generative search should treat repeated querying, source selection, and synthesis as distinct sociotechnical choices that each shape what information becomes visible, credible, and actionable to users.

\section{Related Work}

Our work sits at the intersection of research on online communities as sources of situated knowledge, algorithmic mechanisms that shape visibility within social platforms, and generative
search and AI-mediated information access.

\subsection{Situated Knowledge and Plurality in Online Communities}

People have long turned to online communities for information that is difficult to obtain from institutional sources or from people in their immediate social networks. Work on weak ties shows how distributed social networks provide access to otherwise unavailable information and expertise~\cite{constant1996kindness,granovetter1977strength}, while HCI and CSCW research has documented how people use online communities to seek health information, disclose sensitive experiences, exchange advice, and receive emotional and informational support~\cite{de2014seeking,de2014mental,andalibi2016understanding,andalibi2017sensitive,nambisan2011information}. In these settings, the value of community discourse extends beyond factual answers. Storytelling, self-disclosure, empathy, and accounts of lived experience can provide forms of evidence that complement authoritative information and help others interpret a problem in relation to their own circumstances~\cite{sharma2018mental,10.1145/3290605.3300574}.

Research on advice-seeking further illustrates the importance of this situatedness. \citet{chen202430f} show that self-disclosure can provide a basis for credibility and empathy in Reddit advice exchanges, and can prompt reciprocal self-disclosure from those providing advice. Studies of peer-support communities similarly find that shared experience, trust, recognition, and knowledge exchange are important features of the support these spaces provide \cite{balsamo2023pursuit}. Community members can also contribute in qualitatively different roles---for example, as information providers, emotional-support providers, welcomers, or storytellers---meaning that a discussion thread can contain different \emph{kinds} of useful contributions rather than interchangeable answers~\cite{10.1145/3290605.3300574}.

This heterogeneity also exists in the perspectives communities contain. Members of the same community need not agree about what behavior or outcomes they value, and values can come into conflict within a community~\cite{weld_what_2022,weld2024making}. Across communities, these differences become still more pronounced: subreddits develop distinct norms, rules, expectations, and definitions of desirable behavior~\cite{chandrasekharan2018internet,fiesler2018reddit,weld_what_2022,goyal2026uncovering}. Research showing that the same users quickly adjust their behavior when entering communities with different local norms further demonstrates that community context is not reducible to topic alone~\cite{rajadesingan2020quick}.  Thus, even communities discussing similar problems may differ in what counts as relevant evidence, appropriate advice, or desirable participation. Plurality in online communities therefore exists at multiple levels: among the contributions and experiences within a discussion, among members' values within a community, and across communities with different social contexts. For advice- and support-seeking in particular, encountering these differences can itself be part of the information-seeking process. 

\textit{Our work builds on this literature by asking what happens to such plurality when access to community discourse is mediated by a system that retrieves across communities and condenses their contributions into a single generated answer.}

\subsection{Algorithmic Mediation of Visibility in Social Platforms}

Not every contribution produced within an online community is equally likely to be seen. Social platforms use voting, ranking, recommendation, moderation, and other mechanisms to organize large volumes of user-generated content. These mechanisms interact with human attention and social influence, meaning that observed visibility is itself the product of earlier sociotechnical processes. Experimental work has shown that prior ratings can alter subsequent evaluations and generate cumulative advantage~\cite{muchnik2013social,salganik2006experimental}, while position strongly shapes which content users attend to~\cite{lerman2014leveraging,chan2025ranking}. On Reddit, community feedback is further patterned by the linguistic characteristics of contributions~\cite{goyal2026language,lambert2025does}.

These dynamics have direct implications for which perspectives remain visible. Work on comment-ranking diversification observes that ranking discussion comments primarily using popularity signals can overrepresent majority perspectives and produce redundant top-ranked results, motivating reranking strategies that explicitly incorporate diversity~\cite{northcutt2017comment}.  Moreover, selection need not even be fully algorithmic to have similar consequences. For example, \citet{he2024making}'s study of professionally curated news comments finds systematic differences between selected and non-selected comments illustrating how any curation layer implicitly defines what constitutes a contribution worth foregrounding.

Consequently, popularity or prominence cannot be treated straightforwardly as a measure of either relevance or consensus. A highly visible comment may be prominent partly because earlier users saw and rewarded it, producing path-dependent differences in exposure. This matters for generative search because it introduces an additional selection layer over discourse that has already undergone ranking and social evaluation. A system that preferentially retrieves highly ranked, early, or otherwise visible contributions can inherit and potentially compound these upstream dynamics. 

\textit{Our audit examines this additional layer directly by comparing selected and non-selected material within the same discussions, asking how existing platform visibility, linguistic register, and author standing are associated with whose contributions are carried forward.}

\subsection{Generative Search, Summarization, and Representational Auditing}

Generative search increasingly changes information seeking from navigating a set of documents to receiving an answer synthesized from retrieved evidence~\cite{li2024survey}. Prior evaluations have therefore examined whether generative search systems accurately ground their claims in cited material. \citet{liu2023evaluating} find substantial failures of citation completeness and correctness across answer engines, while subsequent work has identified additional problems involving source selection, citation fidelity, and the relationship between listed sources and generated answers~\cite{venkit2024search,huang2026answer,xu2026measuring,hu2026auditing}. This line of work establishes that the sources appearing behind a generated answer are themselves an important object of study rather than merely supporting metadata.

An emerging HCI literature further asks how generative search changes the information environment users encounter. \citet{10.1145/3613904.3642459} find that LLM-powered conversational search can increase selective exposure relative to conventional search, particularly when the system reinforces a user's existing viewpoint. Other work demonstrates that AI-generated summaries of social-media discussions can alter perceptions of majority opinion and of how balanced the underlying conversation is~\cite{10.1145/3772318.3790945}. Prior work in NLP has also shown that dense retrievers used to retrieve sources before synthesis by a generative model exhibit systematic biases toward short, surface-level sources~\cite{coelho-etal-2024-dwell,fayyaz-etal-2025-collapse,goyal-etal-2026-masking}. These findings highlight a distinction between whether an answer is individually plausible and whether the set of perspectives it presents faithfully represents the underlying information environment.

NLP research on summarization makes this representational problem especially explicit. \citet{zhang-etal-2024-fair} formalize fair summarization of user-generated content in terms of preserving diverse perspectives and find that both LLM-generated and reference summaries can underrepresent perspectives present in their sources. Related work on opinion summarization similarly argues that representing only majority positions can erase minority opinions~\cite{huang-etal-2023-examining}. For community question answering specifically, work on multi-perspective summarization has treated experiences, suggestions, information, and other response types as distinct components worth retaining rather than collapsing them into a single ``best'' answer~\cite{bhattacharya-etal-2022-lchqa}. These concerns are particularly salient for the advice and support settings we study, where the distinction between an individual's experience and a generalized recommendation carries epistemic meaning. At the same time, recent HCI systems have explored how generation might be more faithfully grounded in social context. Social-RAG, for example, retrieves from a group's prior interactions and social signals so that generated content can better reflect that group's preferences and norms~\cite{10.1145/3706598.3713749}. 

Methodologically, we build on the tradition of black-box algorithm auditing, which studies deployed systems through systematic observations of their inputs and outputs when internal mechanisms are unavailable~\cite{sandvig2014auditing,metaxa2021auditing,10.1145/3449148}. Existing generative-search audits have predominantly examined web-scale source selection, factuality, citations, or exposure. Our setting instead treats the retrieval corpus as socially structured. We trace representation across multiple stages---which communities are retrieved, which posts within those communities are recovered, which comments within their discussions are selected, and how those contributions are rewritten during synthesis. We additionally repeat identical queries to distinguish persistent patterns from run-to-run variation. \textit{In doing so, we shift the object of audit from the quality of a generated answer alone to the sociotechnical process through which plural community discourse becomes that answer.}

\section{Data Curation and Community Selection}
\label{sec:data}

For our audit, we wanted to curate queries that actual platform users are likely to ask. We therefore build the query set from the communities themselves, rather than from external question banks, and track provenance from each frozen query back to the post it was distilled from.

\para{Target communities:}
We study twenty subreddits in four groups chosen to span the stakes and epistemics of advice-seeking: \textit{finance}, \textit{support}, \textit{critical advice}, and \textit{life \& decisions}. These communities differ in moderation intensity and in the degree to which questions have factual versus experiential answers. Crucially, each of these communities allows submissions from any member rather than only credentialed ones.

Because community size varies greatly on Reddit, an audit that looked only at the platform's largest communities would speak only to them. Therefore, we deliberately sample \textit{two} sets of ten communities: a set of large, popular communities, and a set of smaller, more niche ones. Set \textsc{Large} (median 3.5M subscribers) covers \textit{finance} (r/personalfinance, r/investingforbeginners), \textit{support} (r/HealthAnxiety, r/mentalhealth, r/techsupport), \textit{critical advice} (r/AskDocs, r/legaladvice), and \textit{life \& decisions} (r/careerguidance, r/relationship\_advice, r/Parenting), and Set \textsc{Small} (median 210K subscribers) mirrors the same structure with r/PersonalFinanceNZ, r/personalfinanceindia, r/MentalHealthUK, r/Anxietyhelp, r/AskTechnology, r/auslegal, r/WomensHealth, r/AskParents, r/Marriage, and r/UKJobs.

Diversifying community sizes matters also because smaller communities are precisely where AI-mediated search might carry the greatest stakes for the community itself, since it has direct consequences on discovery-based newcomers.

\para{Query curation funnel:}

\begin{figure*}[t]
\centering
\includegraphics[width=\linewidth]{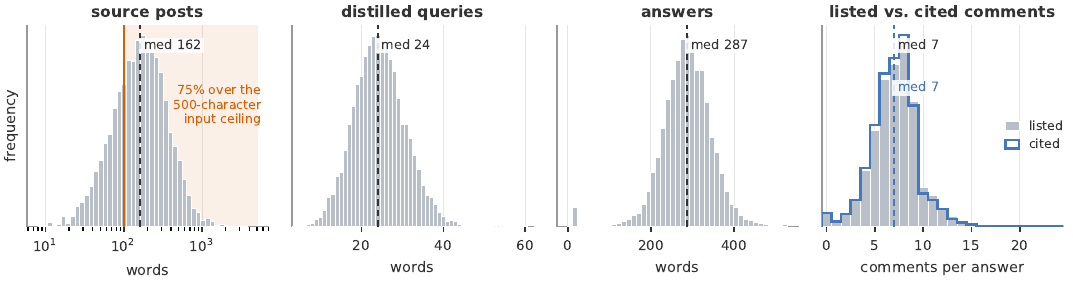}
\caption{\textbf{Length and attribution shape of the audit corpus, both community sets pooled:} Left to right: source-post word counts, distilled queries, synthesized answers, and the number of source comments each answer lists as backend sources vs.\ the subset it quotes inline. The first two panels describe the frozen query set and so count each of the $10{,}000$ queries once; the last two describe answers, which vary by draw, and pool all three draws ($n=30{,}000$). Dashed lines mark medians.}
\Description{Four distribution plots describing the audit corpus. The first shows source-post length on a logarithmic word-count axis, with a median of 162 words and an annotation indicating that roughly 75\% of source posts exceed Reddit Answers' 500-character input ceiling. The second shows the shorter distilled queries, with a median length of 24 words. The third shows generated Reddit Answers responses, with a median length of 287 words. The fourth overlays the number of comments listed as backend sources and the number quoted inline in each answer; both distributions have a median of 7 comments and largely overlap. The query distributions cover 10,000 frozen queries, while the answer-level distributions pool 30,000 answers across three draws.}
\label{fig:corpus}
\end{figure*}

For each subreddit, we curated twelve months of posts (August 2025–July 2026) via the Reddit API\footnote{\href{https://praw.readthedocs.io/en/stable/}{https://praw.readthedocs.io/en/stable/}}, combining multiple listing endpoints (new, top, controversial, and hot) to obtain score- and time-diverse coverage. We initially planned to submit the full posts as search queries, but two considerations led us to curate compact queries instead. First, Reddit Answers enforces a hard input limit of 500 characters and, in practice, returns an answer only for inputs shorter than approximately 100 words. As \autoref{fig:corpus} shows, the original posts have a median length of 162 words, and submitting them verbatim caused the system to refuse roughly 75\% of our sample. Second, and more fundamentally, pasting an entire post does not reflect how people typically query a search system. Prior research shows that user-issued queries tend to be short and keyword-like rather than paragraph-length narratives~\cite{jansen2000real,bendersky2009analysis}. Distilling each post into a standalone question therefore keeps inputs within the system's operating range while better approximating the compressed information need that a user might type into a search interface.

To obtain these distilled queries, we passed each post through a five-stage funnel:
\begin{itemize}
  \item[\textbf{S1.}] \textbf{Pre-filtering:} Remove non-self posts, deleted or removed content, bot accounts, NSFW posts, non-English text, near-empty posts, and
  exact duplicates.

  \item[\textbf{S2.}] \textbf{Stratified sampling:} Sample across within-subreddit score quartiles, so the pool does not over-represent high- or low-score questions.

  \item[\textbf{S3.}] \textbf{Relevance gate:} An LLM (\textit{gpt-5.4-nano}~\cite{openaiGPT54}) keeps only posts expressing a genuine advice-, support-, or information-seeking need. We validate this step with two human annotators on a 100-post sample stratified by source community and model decision (50 \textsc{Keep}, 50 \textsc{Drop}); among posts passed by the LLM, annotators judged 84\% and 88\% to satisfy the inclusion criterion, with full details in Appendix~\ref{app:s3-validation}.

  \item[\textbf{S4.}] \textbf{Query conversion:} An LLM (\textit{gpt-5.4-nano}~\cite{openaiGPT54}) rewrites each kept post into a standalone query while removing explicit demographic markers unless they are essential to the information need. Prior work shows that demographic and role cues can themselves alter retrieval and generated responses~\cite{wu-etal-2025-rag,kaur-etal-2026-whos}, and we return to this choice and its implications in Section~\ref{sec:limitations-demographics}. Since S4 has no unique gold-standard query, we follow prior work using LLMs for document-to-query generation and query rewriting~\cite{bonifacio2022inpars,dai2023promptagator,ma-etal-2023-query}, and report a qualitative manual fidelity audit in Appendix~\ref{app:query-validation}.

  \item[\textbf{S5.}] \textbf{Near-duplicate removal:} Embedding-based deduplication within each subreddit (cosine similarity threshold $\tau = 0.90$) using an embedding model \texttt{Qwen3-Embedding-0.6B}~\cite{zhang2025qwen3}.
\end{itemize}

\noindent From candidate pools of up to 2{,}000 posts per subreddit, we extract exactly 500 frozen queries per subreddit (10{,}000 total). Example queries can be found in \autoref{app:examples}, and prompts issued to LLMs for steps S3 and S4 of the pipeline can be found in \autoref{app:prompts}.

\para{Answers scrape:} We collected each answer by using Reddit Answers in a headless Chromium session. Specifically, for every frozen query we open the Answers search URL and intercept the answers.reddit.com server-sent event (SSE) stream that delivers the response, rather than scraping the rendered page DOM. Parsing that stream yields the synthesized answer text and section structure, inline quote citations with their target comments, the listed source posts, and suggested follow-up questions. Separately, for every listed source post we fetch the full thread---post plus nested comment tree---via the Reddit API. Because the system answers each submission afresh, we repeat the entire scrape three times in immediate succession for both community sets (Section \ref{sec:rq1-variation} explored this in detail), yielding 30{,}000 answers over 337{,}690 threads and 14.68M comments, drawn from 8{,}590 distinct communities. We additionally collected metadata for those communities (subscriber counts, age). See Section \ref{sec:ethics} for Ethical Considerations we kept in mind while scraping data.

\section{RQ1: How Much Do Reddit Answers Vary Across Runs?}\label{sec:rq1-variation}

Because we audit a generative system with several potentially stochastic components—including retrieval of relevant posts and comments, use of the retrieved context, and LLM-based generation—a single scrape cannot be treated as a fixed dataset, but rather as a draw from the system's output distribution. To ensure that the properties we observe characterize the system rather than a particular draw, we directly measure this variation and decompose it by pipeline stage.

\subsection{Methodology}\label{sec:rq1-variation-method}

\subsubsection{Consecutive requery design:} For our study, we chose a consecutive requery design, where we submit the frozen set of 10k queries three times back-to-back, so that consecutive draws are separated by $\sim5$ hours. All passes used identical scrape settings. We chose this design over requerying after days or weeks since that would confound the system's own non-determinism with the drift in the underlying retrieval index as new posts/comments arrive, and total votes shift. However, we believe such an exploration would be a promising direction for future work to derive longitudinal insights.

\subsubsection{Overlap measures:} For each query and pair of draws, we compute the Jaccard overlap of the retrieved post sets, of the comment sets quoted in the Reddit Answer, and of the set of communities drawn from. We also measure the cosine similarity between answer embeddings and the lexical overlap of answer tokens. To make these overlap values more interpretable, we report all measures against the same statistic computed on \textit{mismatched query pairs} drawn from the same community, which gives the overlap expected from topical adjacency alone.

Since our inference target is the property of the output distribution rather than the draw itself, we fit every analysis independently on each draw and combine the resulting estimates for all estimators in the paper using Rubin's rules~\cite{rubin2018multiple}. 

\subsection{Findings}\label{sec:rq1-variation-findings}

\begin{figure*}[t]
\centering
\begin{subfigure}[t]{0.49\textwidth}
  \centering
  \includegraphics[width=\linewidth]{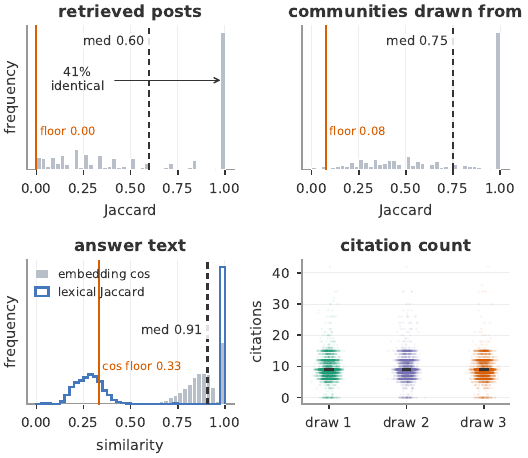}
  \caption{\textsc{Large} communities, 5{,}000 queries.}
  \label{fig:consistency-large}
\end{subfigure}
\hfill
\begin{subfigure}[t]{0.49\textwidth}
  \centering
  \includegraphics[width=\linewidth]{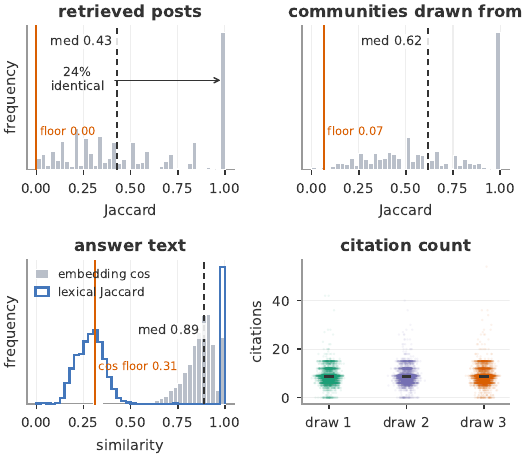}
  \caption{\textsc{Small} communities, 5{,}000 queries.}
  \label{fig:consistency-small}
\end{subfigure}
\caption{\textbf{Agreement between back-to-back re-queries of the same query set, by community set:} Within each panel group: Jaccard overlap of retrieved posts and of communities drawn from, answer-text similarity, and per-query citation counts by draw. Dashed lines mark medians. We find that overlap is sharply bimodal in both sets, with a mode of exactly-identical retrieval and a diffuse mode of weak agreement, but the identical mode is substantially smaller in the smaller communities.}
\Description{Two groups of four plots compare repeated-query stability for 5,000 queries from large communities and 5,000 queries from small communities. For each group, the first plot shows Jaccard overlap of retrieved posts, the second shows Jaccard overlap of communities represented, the third shows similarity between generated answer texts using embedding cosine similarity and lexical Jaccard, and the fourth shows citation counts across the three draws. Retrieval overlap is bimodal, with many exactly repeated source sets and many substantially different ones. Large communities are more reproducible: retrieved-post overlap has median 0.60 and about 41\% of query pairs have identical retrieval, compared with median 0.43 and about 24\% identical retrieval for small communities. Community overlap likewise falls from median 0.75 to 0.62. Despite this retrieval variation, generated answer text remains highly similar, with median embedding cosine similarity of 0.91 for large and 0.89 for small communities, and citation-count distributions remain similar across draws.}
\label{fig:consistency}
\end{figure*}

\subsubsection{Reproducibility shows a bimodal trend:} We find that requerying the same query returns either the same evidence or substantially different evidence, with little in between (see \autoref{fig:consistency-large}). In the \textsc{Large} community set, 41.8\% of pairs retrieve a completely identical source set, and 27.2\% of queries return identical sources in all three draws. Among the pairs that are not identical, agreement is weak with a median Jaccard 0.26 (IQR 0.14--0.44), and 3.7\% of pairs sharing no source post at all. The median overlap is 0.60, against a mismatched-query floor of 0.00. The same patterns hold for the communities drawn from (median 0.75, floor 0.08) and for answer text (embedding cosine 0.91, floor 0.33). For \textsc{Small} communities on the other hand (see \autoref{fig:consistency-small}), we find that these patterns are less reproducible, indicating that reproducibility is not a fixed property of the system but varies with the corpus it is searching. Here, identical retrieval falls from 41.8\% to 24.6\% of pairs, and the share of queries returning identical sources in all three draws falls from 27.2\% to 11.6\%. Median source overlap drops from 0.60 to 0.43, community overlap from 0.75 to 0.62, while answer-text cosine similarity stays similar going from 0.91 to 0.89. The citation count distributions stay stable across runs in both \textsc{Large} and \textsc{Small} sets.

\subsubsection{Most variance is introduced by retrieval:} For \textsc{Large} communities, we find that when two draws retrieve the identical source set, 89.6\% of them return nearly identical answer text (mean embedding cosine $>$0.990). However, when retrieval differs, the same level of identical text occurs only once in 8{,}657 pairs. The \textsc{Small} communities reproduce this finding, with 74.8\% near-identical text given identical retrieval, and 0.01\% given different retrieval. Generation is thus close to deterministic given its inputs, and essentially \textit{all run-to-run variation originates from which documents are retrieved}. This also explains the bimodality we observe in \autoref{fig:consistency}: when the retriever happens to return the same documents, the generator reproduces its previous output almost exactly, and the two modes are simply those of the retriever.

\subsubsection{Refusals are mostly deterministic and silent synthesis is rare:} There are cases where Reddit Answers decides to refuse answering queries. Across the three draws, the refusal rate is stable at 1.06\%, 1.06\%, and 1.04\% (unanimous on 99.5\% of queries, Fleiss $\kappa = 0.83$), confirming refusal is a property of the query. Some examples of refusal answers include \textit{``Reddit doesn't provide responses to some prompts, including those that are potentially unsafe or may be in violation of Reddit's policies,''} \textit{``As an AI, I don't have personal experiences or a physical body, so I can't go to the ER or experience anxiety,''} and \textit{``I am Reddit Answers, a language model. I don't have personal experiences or make investments. I can, however, search Reddit for discussions... .''} We also find that silent synthesis, i.e., listing sources but quoting nothing inline, occurs at rates 0.20\%, 0.24\%, and 0.22\% ($\kappa = 0.45$), which is rare, and nearly independent across draws. We therefore attribute this to transient generation failure of the system.

\begin{takeawaybox}
\textbf{\faLightbulbO ~\large Takeaway:}
Repeated queries reveal that Reddit Answers is not a fixed system but a distribution over possible source sets and answers. Most run-to-run variation is introduced by retrieval rather than generation, and this instability is substantially greater for smaller communities.
\end{takeawaybox}

\section{RQ2: Which Communities Do Answers Draw From?}\label{sec:rq2}

Reddit cannot be considered a single corpus, and is made up of thousands of communities each with its own population and accumulated expertise, while sharing norms and values~\cite{chandrasekharan2018internet,goyal2026uncovering,weld_what_2022}. A question about a medical diagnosis can be answered using a clinical community, a patient-support community, or a general chat community, and the three may not answer the question in the same way or with the same nuance. Therefore when Reddit Answers synthesizes a response, it makes that choice on the user's behalf and the answer surfaces perspectives of the selected communities. In RQ2, we ask how that choice is made and what it systematically favors (see \autoref{fig:reddit-answers-pipeline} \textbf{RQ2}).

Since every query in our dataset is distilled from a real post, we always know one community in which this exact question was already asked and answered, along with the specific post it was asked in. This lets us use this post as a reference point to ask both \textit{(1) which communities appear most often}; and \textit{(2) whether the system reaches the original post, and if not, whether what it finds is any closer to answering the question.} 

\subsection{Methodology}
\label{sec:rq2-method}

\subsubsection{Characterizing cross-community routing:}
Because every query in our dataset is linked to the community from which its source post was drawn, we can use that community as a consistent reference point for characterizing where retrieval goes. For each query $q$ with source community $c(q)$, we compute the \textit{home share}, defined as the fraction of retrieved posts whose threads also belong to $c(q)$. We do not treat higher home share as inherently better. Instead, it measures how often retrieval remains within versus crosses the boundary of the source community. For the posts retrieved elsewhere, we characterize the size of their communities using subscriber counts.

\subsubsection{Using the origin post as a known-relevant retrieval probe:}
Our query construction also gives us one post that we know is directly relevant to each query. Specifically, every query was distilled from an origin post $o$ that had already expressed the underlying information need. We therefore use retrieval of this post as a probe of whether the system recovers a known-relevant document, rather than as a claim that the origin is the uniquely correct source. Let $\mathrm{origin}_q$ indicate whether the origin post appears among the answer's listed sources. We model
\begin{equation}
\label{eq:m3}
\mathrm{logit}\ \Pr(\mathrm{origin}_q = 1)
=
\theta_0
+
\theta_1 \log(1+S^{o}_{q})
+
\boldsymbol{\psi}^{\top}\mathbf{C}_q,
\end{equation}
where $S^{o}_{q}$ is the origin post's score and $\mathbf{C}_q$ contains source-community fixed effects. This lets us test whether community-conferred visibility predicts whether an otherwise known-relevant post is recovered.

\subsubsection{Measuring within-community substitutability:}
Whether the origin post is retrieved may also depend on how distinctive it is within its own community. If a community contains many posts asking nearly the same question, the retriever may have several plausible substitutes and no single post may stand out. We measure this from each community's twelve-month post pool $P_c$ by embedding queries and posts with \texttt{Qwen3-Embedding-0.6B}. For a query $q$ distilled from origin post $o$, we compute
\begin{equation}
\label{eq:subst}
\begin{split}
n^{\mathrm{asgood}}_q
&=
\bigl|\{\,p \in P_c \setminus \{o\} :
\cos(q,p) \ge \cos(q,o)\,\}\bigr|, \\[2pt]
\mathrm{margin}_q
&=
\cos(q,o)
-
\max_{p \in P_c \setminus \{o\}}
\cos(q,p).
\end{split}
\end{equation}
The first quantity counts how many in-community posts match the query at least as well as its origin, while the second measures how far the origin stands above its closest alternative. Both are standardized before entering the model in Eq.~\ref{eq:m3}, so their coefficients are interpreted per standard deviation.

To capture the same idea at the community level, we additionally compute \textit{homogeneity}, defined as the mean pairwise cosine similarity among posts in $P_c$. Higher values indicate communities whose discussions are more interchangeable in semantic space. Because homogeneity is constant within a community and therefore collinear with the community fixed effects in Eq.~\ref{eq:m3}, we estimate its relationship with retrieval only at the community level. These substitutability measures depend on the fixed query set rather than on any particular scrape, so they are computed once per cohort.

\subsubsection{Testing the semantic fit of retrieved posts:}
The routing measures above tell us where retrieval lands, but not whether leaving the source community leads to semantically better evidence. We test this directly on a random subsample of 589 queries by embedding the query, its origin post, and every post retrieved for it with \texttt{Qwen3-Embedding-0.6B}. We then compare the query similarity of each retrieved post, $\cos(q,p)$, against the similarity of the known-relevant origin, $\cos(q,o)$. This gives us a simple reference point for interpreting cross-community movement. If retrieved posts are consistently less similar to the query than the origin post the system passed over, then routing away from the source community cannot be explained simply by finding a better semantic match elsewhere.

\subsection{Findings}
\label{sec:rq2-findings}

\begin{figure*}[t]
    \centering
    \includegraphics[width=\linewidth]{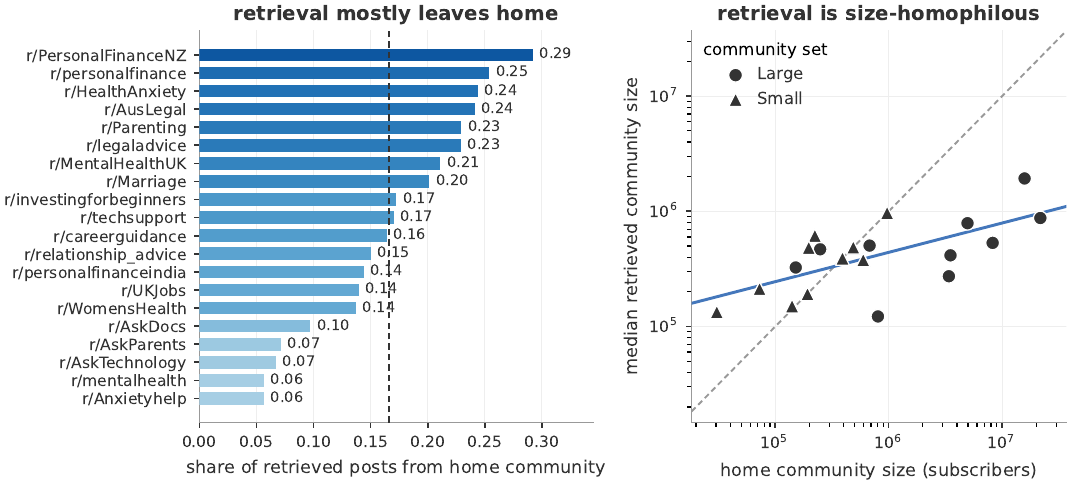}
    \caption{\textbf{(left)} \textbf{Share of retrieved posts drawn from the query's source community:} The dashed line marks the overall mean.
    \textbf{(right)} \textbf{Median size of the communities retrieval lands in against the size of the source community:} The fitted line shows the size-homophily relationship and the dashed diagonal marks equality.}
    \Description{Two plots characterize where Reddit Answers retrieves evidence across communities. The left horizontal bar chart shows, for each of 20 source subreddits, the fraction of retrieved posts that come from the same community as the original query. Home-community shares are low for every subreddit, ranging from about 0.06 to 0.29, indicating that retrieval usually crosses community boundaries. The right scatterplot places source-community subscriber count on the horizontal axis and median subscriber count of retrieved communities on the vertical axis, both on logarithmic scales. Points distinguish the large and small community sets. The positive fitted trend indicates size homophily: larger source communities tend to retrieve from larger communities. Relative to the equality diagonal, large communities often retrieve from smaller communities, while small communities frequently retrieve from substantially larger ones.}
    \label{fig:routing}
\end{figure*}

\subsubsection{Reddit Answers assembles answers across communities:}
Rather than drawing repeatedly from one part of Reddit, Reddit Answers typically assembles each response from a broad set of communities and posts. A typical answer spans roughly seven distinct communities in both the \textsc{Large} and \textsc{Small} sets. Only 17.7\% of retrieved posts in the \textsc{Large} set and 15.7\% in the \textsc{Small} set come from the community where the query originated. We do not treat staying within this ``home'' community as an ideal the system should maximize. Instead, because every query has a known source community, home share gives us a consistent reference point for measuring how extensively retrieval crosses community boundaries. As shown in Figure~\ref{fig:routing}, this cross-community routing occurs for every community we study. Retrieval is also broadly dispersed. The most-reused 1\% of source posts appear in only 5.1\% of answers in the \textsc{Large} set and 3.5\% in the \textsc{Small} set, so the system is not simply relying on a small set of canonical threads.

\subsubsection{Cross-community routing follows community size:}
The communities that enter an answer are not selected independently of where the query came from. Across our twenty communities, the size of the source community predicts the size of the communities retrieval reaches ($\beta = +0.21$ per $\log_{10}$ subscriber, $p=0.007$). A tenfold larger source community draws from communities roughly $1.6\times$ larger. The direction of this movement differs across our two sets. In the \textsc{Large} set, 61.8\% of retrieved posts come from smaller communities, with retrieval often moving toward more specialized spaces. In the \textsc{Small} set, the pattern reverses. Roughly two-thirds of retrieved posts come from larger communities. Questions originating in smaller and potentially more niche spaces are therefore often answered using discourse from substantially larger ones.

\subsubsection{Already-visible questions are easier to recover:}
The origin posts give us a second way to characterize which material the system is more likely to retrieve. Each origin post is a document that we know expressed the query before it was distilled, so it provides one directly relevant item that the system could recover. Reddit Answers retrieves this post for 35.7\% of queries in the \textsc{Large} set and 42.6\% in the \textsc{Small} set, and this outcome is highly stable across draws ($\kappa=0.87$ and $0.85$). More importantly, recovery of the original post is correlated with community-conferred visibility. Each $e$-fold increase in the origin post's score raises its retrieval odds by 24.1\% in the \textsc{Large} set (OR 1.241, [1.205, 1.279]) and 16.4\% in the \textsc{Small} set (OR 1.164, [1.118, 1.211]). Even among posts that we know are directly relevant, those already amplified by the community are easier for the system to find.

\subsubsection{Redundancy predicts which posts are recovered:}
Beyond visibility, retrieval also depends on how distinctive a post is within its own community. When a community contains many other posts that match a query just as well as its origin, the origin is less likely to be retrieved (OR 0.851 in the \textsc{Large} set and 0.842 in the \textsc{Small} set per standard deviation). Conversely, posts that stand out as the clearest match within their community are much easier to retrieve, with a one-SD increase in their advantage over the next-best match roughly doubling retrieval odds (OR 2.044 and 1.869). The same pattern appears at the community level. Mean within-community homogeneity strongly predicts the closed-loop rate ($\beta=-2.03$, $p<10^{-4}$, $R^2=0.68$), even after controlling for community size. In the \textsc{Small} set, adding the query-level redundancy measures explains 33.8\% of the spread between community fixed effects. Communities that repeatedly discuss similar problems therefore give the retriever many near-substitutes, making any particular contribution harder to recover.

\subsubsection{Cross-community routing is not explained by better semantic matches:}
To understand whether this cross-community movement is simply helping the system find better evidence, we compare retrieved posts against the one post we already know is directly relevant. On our semantic-fit subsample, the origin post has an average query similarity of 0.79, compared with 0.60 for retrieved posts and 0.72 even for the single best retrieved post. Only 5.0\% of retrieved posts are closer to the query than its origin, and for 79.3\% of queries not one retrieved post is a closer match. Retrieved posts from the source community are also marginally closer than those retrieved elsewhere. The extensive cross-community routing we observe therefore cannot be explained simply as the system leaving one community to find semantically better matching material.

\begin{takeawaybox}
\textbf{\faLightbulbO ~\large Takeaway:}
Reddit Answers constructs each response by assembling evidence across communities, but this routing is systematic rather than neutral. Which communities and posts enter an answer is associated with community size, prior visibility, and how distinctive a post is within its local context. Moreover, moving across communities is usually not explained by finding semantically better matching material.
\end{takeawaybox}

\section{RQ3: Whose Comments Become the Answer?}\label{sec:rq3}

Having established which communities retrieval draws from, RQ3 asks which comments Reddit Answers selects from the threads it decides to use. We break this question into three parts, focusing on \textbf{(i) thread structure:} the platform-visible signals a ranker could use without reading the comment, \textbf{(ii) language:} the register in which a comment is written, and \textbf{(iii) author:} the standing of the person who wrote it.

\subsection{Methodology}
\label{sec:rq3-method}

\subsubsection{Defining comment selection:}
Reddit Answers exposes two closely related selection outcomes. It \textit{lists} comments as backend sources and \textit{quotes} some of these as inline sources. Across all six draws, 97.2--97.7\% of listed comments are also quoted. Rather than treating these as separate decisions, we therefore define a single outcome,
$\mathrm{Sel}_{jt}=\mathbf{1}[\text{comment } j \text{ is listed or quoted}]$,
and report the listed-versus-quoted contrast only as an exploratory observation. Our unit of analysis is a comment $j$ in thread $t$.

\subsubsection{Comparing comments within the same thread:}
All comment-level models are conditional logistic regressions with a thread fixed effect $\alpha_t$. This absorbs thread-level differences such as topic, popularity, community, query, and time, so coefficients are identified from contrasts among comments appearing within the same thread~\cite{chamberlain1980analysis}. Threads without outcome variation do not contribute to the conditional likelihood, leaving approximately 20{,}600--21{,}800 threads per draw with at least one selected comment. Within large threads, non-selected comments are uniformly subsampled with a cap of 30 comments per thread, which preserves consistency of the conditional-logit estimates. In addition to effect sizes, we control the false discovery rate using Benjamini--Hochberg correction~\cite{benjamini1995controlling}.

\subsubsection{Modeling thread structure and metadata:}
Our first feature set captures the structural and platform-visible properties of each comment:
\begin{equation}
\label{eq:m1}
\begin{split}
\mathrm{logit}\ \Pr(\mathrm{Sel}_{jt}=1) = \alpha_t
&+ \beta_1\,\mathrm{ScorePct}_{jt} + \beta_2\,\mathrm{TopLevel}_{jt}\\
&+ \beta_3\log(1{+}D_{jt}) + \beta_4\log(1{+}W_{jt}) \\
&+ \beta_5\log(1{+}A_{jt}) + \beta_6\,\mathrm{IsOP}_{jt}
+ \boldsymbol{\gamma}^{\top}\mathbf{Z}_{jt},
\end{split}
\end{equation}
where $\mathrm{ScorePct}\in[0,1]$ is the comment's within-thread score percentile, $D$ is its reply depth, $W$ its length in words, and $A$ its age in hours relative to the post. $\mathbf{Z}$ contains indicators for whether the comment includes a link, is marked controversial, is stickied, or is moderator-distinguished.

\subsubsection{Modeling language and register:}
Our second feature set captures how a comment is written using a combination of Linguistic Inquiry and Word Count (LIWC-15)~\cite{pennebaker2015development} and transformer-based measures. We construct three composite scores from the means of $z$-scored categories: \textit{experiential voice} (\texttt{i}, \texttt{ppron}, \texttt{focuspast}, \texttt{feel}, \texttt{percept}), \textit{negative affect} (\texttt{negemo}, \texttt{anx}, \texttt{sad}, \texttt{anger}), and \textit{prosocial warmth} (\texttt{social}, \texttt{affiliation}, \texttt{friend}, \texttt{family}). We additionally include positive affect, hedging, certainty, specificity, and \textit{directive prescription}, measured through obligation modals such as \textit{should} and \textit{must}, along with classifier-based measures of \textit{formality}~\cite{dementieva-etal-2023-detecting} and \textit{toxicity}~\cite{Detoxify}. These linguistic features have been widely used to study advice seeking and support in online communities~\cite{de2014mental,sharma2018mental,de2017language}, social media discourse more broadly~\cite{tausczik2010psychological,goel2026ai,goyal2026language,zhou2026ai}, and generative search outputs~\cite{huang2026answer,venkit2024search,saha2026ai,saha2026linguistic}.

\subsubsection{Modeling author reputation:}
Our final feature set asks whether author standing is associated with selection beyond the content and position of the comment. We add log(\textit{karma}), account age, moderator status, and premium status to Eq.~\ref{eq:m1} on a 1:1 within-thread case-control sample containing every selected comment's author and one non-selected author from the same thread~\cite{vandenbroucke2012case,shimgekar2026detecting}. Outcome-based sampling preserves the logistic coefficients, while drawing controls within threads maintains the fixed-effects design. Karma captures accumulated platform reputation through prior votes, moderator status captures an institutional role within the community, and premium status captures whether the account has Reddit Premium.

\subsubsection{Separating overall and visibility-adjusted associations:}
The linguistic and author features above are not independent of the visibility signals in our structural model. A comment's language and author standing may themselves be associated with the votes and position it receives. Conditioning on these variables therefore changes the quantity being estimated and, under a causal interpretation, could also condition on mediators or open collider paths~\cite{dafoe2015confounding,pearl2009causality}. We therefore report two specifications for each feature. The \textbf{overall association} controls only for comment length, since rate-based linguistic features are mechanically length-sensitive. The \textbf{visibility-adjusted association} additionally controls for $\mathrm{ScorePct}$, $\mathrm{TopLevel}$, and $\log(1{+}A)$. Because logit coefficients cannot be directly compared across nested specifications~\cite{mood2010logistic}, we use the Karlson--Holm--Breen (KHB) method~\cite{karlson2012comparing} to quantify how much the association changes after accounting for these visibility variables. We note that these quantities remain observational and should not be interpreted as identified causal direct or indirect effects.

\subsubsection{Comparing what the community rewards with what the system selects:}
Prior work shows that upvotes are themselves patterned by how comments are written and by the behaviors communities reward~\cite{goyal2026language,lambert2025does}. A linguistic feature may therefore be associated with both community reception and Reddit Answers selection. To compare these two processes, we additionally regress within-thread score percentile on the same ten linguistic features using thread fixed effects and standard errors clustered by thread. This lets us compare whether community voting and system selection favor the same kinds of language. Because the two models have different outcome scales, we compare the direction and pattern of associations rather than their coefficient magnitudes.

\subsubsection{Robustness to unmeasured answer relevance:}
A remaining concern is that comments may be selected simply because they more directly answer the query, an unobserved property that may itself correlate with visibility, structure, and linguistic register. We therefore replicate our analyses among top-level comments only and use Oster bounds~\cite{oster2019unobservable} to assess sensitivity to remaining unmeasured confounding in Appendix~\ref{app:rq3-robustness}. Our central findings on visibility, formality, and experiential voice remain robust, while several smaller associations do not, and we avoid relying on those unstable effects in our main interpretation.

\subsection{Findings}
\label{sec:rq3-findings}

\begin{figure*}[t]
\centering
\includegraphics[width=\linewidth]{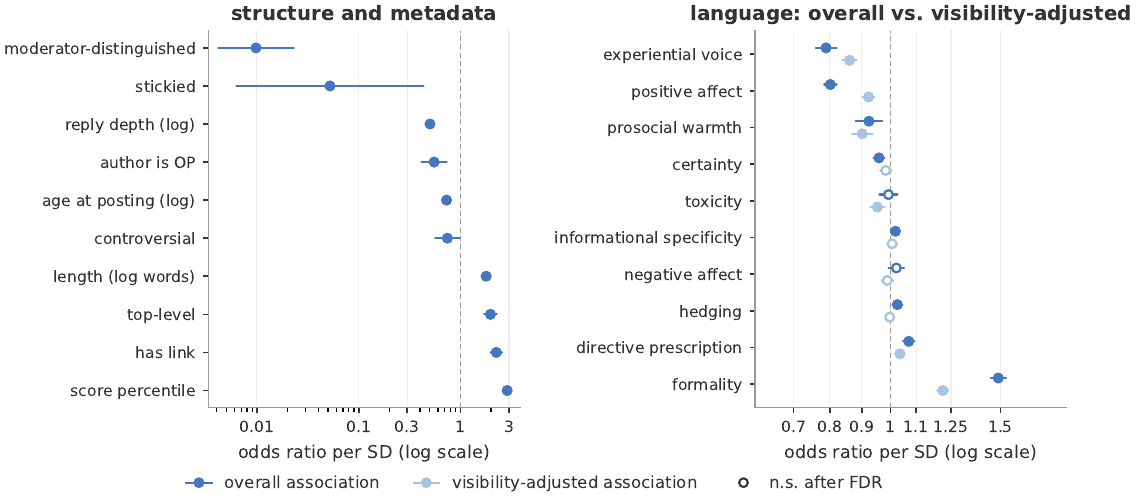}
\caption{\textbf{Selection odds ratios per standard deviation (log scale, 95\% CIs), Rubin-combined over six draws:} \textbf{Left:} structure and metadata (Eq.~\ref{eq:m1}). \textbf{Right:} language, shown as overall associations (dark) and visibility-adjusted associations (light). Filled markers are significant after Benjamini--Hochberg correction; hollow markers are not.}
\Description{Two forest plots show odds ratios for whether a comment is selected by Reddit Answers, with a vertical reference line at an odds ratio of 1 and 95\% confidence intervals. The left plot covers thread structure and metadata. Higher within-thread score percentile is the strongest positive predictor, followed by having a link, being top-level, and greater comment length. Greater reply depth and later posting reduce selection odds, while moderator-distinguished and stickied comments have especially low selection odds. The right plot covers linguistic features and compares overall associations with associations adjusted for visibility. Formality has the strongest positive association with selection and remains positive after adjustment; directive language is also favored. Experiential voice and positive affect have the clearest negative associations and remain negatively associated after accounting for visibility. Several smaller language effects are near an odds ratio of 1 or are not statistically significant after correction.}
\label{fig:forest}
\end{figure*}

\begin{figure}[t]
\centering
\includegraphics[width=0.5\linewidth]{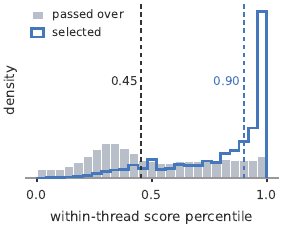}
\caption{\textbf{Distribution of within-thread score percentile for selected vs.\ unselected comments.} Selected comments sit at a median percentile of 0.91, against 0.45 for comments the system passed over.}
\Description{Overlaid distributions of within-thread score percentile for comments selected by Reddit Answers and comments passed over by the system. Passed-over comments are broadly distributed across the percentile range, with a median score percentile of 0.45. Selected comments are heavily concentrated near the top of the score distribution, with a sharp peak close to the 100th percentile and a median of 0.90. Vertical dashed lines mark the two medians, showing that selected comments had substantially greater community-conferred visibility than comments that were not selected.}
\label{fig:scorepct}
\end{figure}

\subsubsection{The system picks what was already visible:}
Thread structure and platform-visible signals strongly predict whether a comment appears in Reddit Answers (Figure~\ref{fig:forest}, left). A one-standard-deviation increase in within-thread score percentile multiplies the odds of selection by 2.88 (95\% CI [2.70, 3.07], $q<10^{-4}$), the largest association in the model. Being a top-level comment nearly doubles the odds of selection (OR 1.97, [1.71, 2.28], $q<10^{-4}$), while greater reply depth sharply reduces them (OR 0.50 per SD, [0.46, 0.55], $q<10^{-4}$). Descriptively, selected comments sit at a median score percentile of 0.91 compared with 0.45 for those passed over (Figure~\ref{fig:scorepct}), and 92\% are top-level compared with 53\% of all comments. Earlier comments are also favored (OR 0.73 per SD of log age, [0.71, 0.75], $q<10^{-4}$), arriving a median 1.2 hours after the post compared with 5.9 hours for the remainder.

These patterns suggest that the first-mover dynamics known to shape vote accumulation~\cite{muchnik2013social,lerman2014leveraging} carry forward into the Reddit Answers selection layer. Longer comments are more likely to be selected (OR 1.79, [1.72, 1.86], $q<10^{-4}$), as are comments containing external links (OR 2.25, [1.99, 2.55]). Moderator-distinguished comments (OR 0.010, [0.004, 0.023], $q<10^{-4}$) and stickied comments (OR 0.052, [0.006, 0.434], $q=0.007$) are almost entirely excluded, while the OP's own comments are also disfavored (OR 0.55, [0.42, 0.73], $q<10^{-4}$). Reddit Answers therefore does not sample the discussion uniformly. It disproportionately surfaces comments that were already prominent within the thread.

\subsubsection{The system prefers prose that already sounds like an answer:}
Language also strongly differentiates what Reddit Answers selects. Among our ten linguistic predictors, \textit{formality} is the strongest, with a one-SD increase raising selection odds by 49\% (OR 1.488, [1.447, 1.531], $q<10^{-4}$). \textit{Directive prescription} also increases selection odds (OR 1.070, [1.048, 1.093], $q<10^{-4}$), while \textit{informational specificity} has a smaller association (OR 1.019, $q=0.045$). In contrast, \textit{experiential voice} (OR 0.789, [0.759, 0.820], $q<10^{-4}$), \textit{positive affect} (OR 0.801, [0.784, 0.820], $q<10^{-4}$), and \textit{prosocial warmth} (OR 0.924, [0.880, 0.970], $q=0.003$) are all associated with lower selection odds. Negative affect and toxicity show no overall association.

Within established typologies of peer support, this amounts to a systematic shift in the kinds of contributions that survive selection. Markers of nondirective support such as personal experience, warmth, and encouragement are disfavored, while formal and directive language is rewarded~\cite{prescott2017peer}. Reddit Answers therefore does not merely compress the conversation. It preferentially selects one recognizable mode of helping over another.

This preference for answer-like prose is only partly accounted for by community visibility. For \textit{experiential voice}, the KHB decomposition attributes 37\% of the overall association to score and position, while a substantial visibility-adjusted association remains (OR 0.860, [0.838, 0.883], $q<10^{-4}$). The AI selection layer therefore disfavors first-person testimony beyond what is captured by the community's own amplification signals. Formality shows a similar pattern, with 51\% of the association accounted for by visibility and a remaining adjusted OR of 1.213 ([1.189, 1.238]). Positive affect differs, with 64\% of its association accounted for by the visibility variables and a smaller adjusted association remaining (OR 0.923).

\subsubsection{The system's preferences are not identical to the community's:}
Comparing what predicts voting with what predicts system selection shows considerable agreement in direction, but also several meaningful differences. Formality is favored by both community voting and Reddit Answers selection, while experiential voice and positive affect are disfavored in both. Because these outcomes are measured on different scales, we do not compare the magnitudes of their coefficients directly.

The comparison also clarifies two apparent null results. \textit{Toxicity} and \textit{negative affect} show no overall association with selection (OR 0.993 and 1.022, both n.s.). Yet community voting is positively associated with both ($\beta=+0.011$, $q=0.005$ and $\beta=+0.016$, $q<10^{-14}$), while toxicity becomes negatively associated with selection after visibility is held fixed (OR 0.953, $q<10^{-3}$). What initially appears to be indifference therefore masks opposing patterns between community amplification and the subsequent AI selection layer.

\subsubsection{Contested and downvoted comments rarely survive selection:}
The visibility preference is especially pronounced for comments that are unpopular or contested. Only 0.53\% of selected comments have a non-positive score, compared with 5.1\% of all fetched comments, a roughly tenfold under-representation. Descriptively, selected comments are also about one-sixth as likely to carry Reddit's \textit{controversial} flag. Within our conditional model, controversial comments remain less likely to be selected (OR 0.744, [0.563, 0.983], $q=0.038$). Downvoted and disputed perspectives can therefore surface, but they are substantially less likely to survive the selection stage. This makes minority or contrarian positions less visible before synthesis even begins.

\subsubsection{Author reputation operates largely through existing visibility:}
Author-level signals provide a different pattern. Karma has no significant overall association with selection (OR 1.071, [0.981, 1.170] per SD, $q=0.125$), but the KHB decomposition identifies a strongly positive visibility-associated component. This is consistent with higher-karma authors having greater community visibility, which itself predicts selection. Account age follows a similar but smaller pattern (overall OR 1.059, [1.030, 1.089], $q<10^{-3}$, with 21\% accounted for by visibility). Descriptively, selected authors hold roughly twice the karma of their same-thread counterparts, but much of this difference is accounted for by the visibility their comments have already accumulated.

Two author attributes depart from this pattern. Moderators are selected less often (overall OR 0.783, [0.726, 0.844], $q<10^{-4}$; adjusted OR 0.866), while Reddit Premium holders are also strongly disfavored (overall OR 0.280, [0.198, 0.395], $q<10^{-4}$; adjusted OR 0.257). Author reputation therefore matters less as a direct selection signal than the visibility that platform standing has already helped produce.

\begin{takeawaybox}
\textbf{\faLightbulbO ~\large Takeaway:}
Reddit Answers adds another layer of ranking on top of the community's own. It disproportionately selects comments that are already visible, early, and top-level, while favoring formal and directive prose over experiential and supportive language. These preferences are only partly explained by community voting, and contested or downvoted perspectives are especially unlikely to survive selection.
\end{takeawaybox}

\section{RQ4: How Does Synthesis Transform the Community's Voice?}
\label{sec:rq4}

While RQ2 and RQ3 characterize what enters an answer, RQ4 asks what happens to those contributions once they are synthesized. Reddit Answers does not simply reproduce the comments it retrieves. It rewrites them into a unified response, potentially changing how the community's contributions sound by the time they reach the user. Recent surveys and HCI research further suggest that people increasingly turn to general-purpose LLMs for advice- and information-seeking activities~\cite{chatterji2025people,10.1145/3772318.3791050,10.1145/3772318.3791647,zhou2026ai}. We therefore compare Reddit Answers against off-the-shelf web-grounded LLMs to distinguish changes specific to community-grounded synthesis from those that may characterize generative answering more broadly.

\subsection{Methodology}
\label{sec:rq4-method}

\subsubsection{Comparing register across community and generated responses:}
For each query, we compare three forms of text: the community comments quoted by Reddit Answers, the Reddit Answers response synthesized from them, and an answer to the same query from a general-purpose LLM. We score all three using the same LIWC- and transformer-based measures introduced in RQ3. Comparing Reddit Answers against its quoted source comments captures how synthesis transforms the material it directly draws from, while comparison with general-purpose LLMs helps distinguish Reddit-specific transformations from broader patterns in generative answering. Every system answers the same set of 1{,}000 queries, stratified to include 50 queries from each community, and we report paired differences and standardized effect sizes using Cohen's $d$.

\subsubsection{Comparing where general-purpose LLMs seek evidence:}
We submit the same queries to two general-purpose LLMs, \textit{GPT-4o-mini}~\cite{hellogpt4o} (\texttt{gpt-4o-mini-2024-07-18}) and \textit{GPT-5}~\cite{openaiIntroducingGPT5} (\texttt{gpt-5-2025-08-07}), through the OpenAI API with web search available. We use two conditions. In the \textit{natural} condition, the model decides whether to search before answering. In the \textit{forced} condition, the model must search before producing its response. The natural condition captures whether general-purpose systems seek external evidence for these advice-seeking queries at all, while the forced condition allows us to compare which sources they select when search is held constant. For every answer, we record whether search was used and the domain of every cited source.

We additionally audit whether individual claims in Reddit Answers are supported by the comments attributed to them, using a two-verifier analysis described in Appendix~\ref{app:rq4-support}. While verifier disagreement prevents us from treating either unsupported-claim rate as a precise estimate, both methods identify cases where generated claims cannot be traced to the evidence Reddit Answers attributes to them.

\providecolor{bardown}{HTML}{2166AC}
\providecolor{barup}{HTML}{762A83}
\providecommand{\dbar}[3]{%
  \makebox[0.42cm][r]{\textcolor{bardown}{\rule[0.45ex]{#1}{4.5pt}}}%
  \makebox[0.60cm][c]{\scriptsize #2}%
  \makebox[0.42cm][l]{\textcolor{barup}{\rule[0.45ex]{#3}{4.5pt}}}}
\begin{table*}[t]
\centering\small\sffamily
\caption{\textbf{RQ4. How each system's answer differs in register from the community comments:} Every system answers the same queries, so all comparisons are paired at the query level: $t$ is a paired $t$-test against the community text (\textsuperscript{*}$q<0.05$, \textsuperscript{**}$q<0.01$, \textsuperscript{***}$q<0.001$), Benjamini--Hochberg corrected within each system, and $d$ is Cohen's $d$, shaded \textcolor{poscolor}{\textbf{teal}} when positive and \textcolor{negcolor}{\textbf{orange}} when negative. Bar length encodes the magnitude of the relative change, \textcolor{bardown}{\textbf{blue}} for a \textbf{decrease} and \textcolor{barup}{\textbf{purple}} for an \textbf{increase}.}
\Description{Table comparing the linguistic register of community comments with answers from Reddit Answers, GPT-4o-mini with forced search, and GPT-5 with forced search across four groups of measures: voice and testimony, register and stance, affect and tone, and social and instrumental language. Values report each community baseline mean, each system mean, percent difference, paired t-test, and Cohen's d. All three systems substantially reduce markers of personal testimony: first-person singular language decreases by 98 percent for Reddit Answers, 92 percent for GPT-4o-mini, and 72 percent for GPT-5; personal pronouns and past-tense focus also decrease strongly across all systems. Reddit Answers and GPT-4o-mini become more formal than community comments, whereas GPT-5 becomes less formal. All systems sharply reduce toxicity and swearing. Causal-reasoning and work-related language increase across all three systems, while several other affective and social measures vary by system. Overall, the most consistent cross-system transformation is a weakening of first-person experiential voice, with Reddit Answers showing the largest reduction.}
\label{tab:rq4-register}
\resizebox{\linewidth}{!}{
\begin{tabular}{@{}l r r c r r r c r r r c r r@{}}
 & \textbf{Community} & \multicolumn{4}{c}{\textbf{Reddit Answers}} & \multicolumn{4}{c}{\textbf{GPT-4o-mini (Forced Search)}} & \multicolumn{4}{c}{\textbf{GPT-5 (Forced Search)}} \\
\cmidrule(lr){3-6}\cmidrule(lr){7-10}\cmidrule(lr){11-14}
\textbf{Metric} & mean & mean & Diff. \% & $t$ & $d$ & mean & Diff. \% & $t$ & $d$ & mean & Diff. \% & $t$ & $d$ \\
\midrule%
\rowcolor{blue!10}\multicolumn{14}{@{}l}{\textit{\textbf{Voice and testimony}}} \\[1pt]
First-person singular & 0.033 & 5.7E$-$4 & \dbar{0.413cm}{-98\%}{0.000cm} & -43.9*** & \gradcell{-1.93} & 0.003 & \dbar{0.385cm}{-92\%}{0.000cm} & -40.7*** & \gradcell{-1.77} & 0.009 & \dbar{0.303cm}{-72\%}{0.000cm} & -31.1*** & \gradcell{-1.32} \\
\rowcolor{gray!10}Personal pronouns & 0.095 & 0.038 & \dbar{0.254cm}{-61\%}{0.000cm} & -63.2*** & \gradcell{-1.98} & 0.041 & \dbar{0.238cm}{-57\%}{0.000cm} & -58.1*** & \gradcell{-1.93} & 0.048 & \dbar{0.210cm}{-50\%}{0.000cm} & -54.8*** & \gradcell{-1.66} \\
Past-tense focus & 0.027 & 0.009 & \dbar{0.287cm}{-68\%}{0.000cm} & -36.0*** & \gradcell{-1.43} & 0.008 & \dbar{0.302cm}{-72\%}{0.000cm} & -37.7*** & \gradcell{-1.58} & 0.009 & \dbar{0.274cm}{-65\%}{0.000cm} & -33.7*** & \gradcell{-1.41} \\
\rowcolor{gray!10}Feeling & 0.007 & 0.007 & \dbar{0.000cm}{+9\%}{0.037cm} & 2.3* & \gradcell{+0.07} & 0.006 & \dbar{0.016cm}{-4\%}{0.000cm} & -0.9 & \gradcell{-0.03} & 0.008 & \dbar{0.000cm}{+25\%}{0.104cm} & 6.4*** & \gradcell{+0.20} \\
Perceptual & 0.020 & 0.019 & \dbar{0.015cm}{-3\%}{0.000cm} & -1.4 & \gradcell{-0.04} & 0.018 & \dbar{0.046cm}{-11\%}{0.000cm} & -4.5*** & \gradcell{-0.14} & 0.021 & \dbar{0.000cm}{+5\%}{0.021cm} & 2.1* & \gradcell{+0.07} \\
\addlinespace[2pt]
\rowcolor{blue!10}\multicolumn{14}{@{}l}{\textit{\textbf{Register and stance}}} \\[1pt]
Formality & 0.652 & 0.821 & \dbar{0.000cm}{+26\%}{0.109cm} & 20.3*** & \gradcell{+0.80} & 0.902 & \dbar{0.000cm}{+38\%}{0.161cm} & 32.6*** & \gradcell{+1.31} & 0.537 & \dbar{0.074cm}{-18\%}{0.000cm} & -14.7*** & \gradcell{-0.58} \\
\rowcolor{gray!10}Hedging & 0.040 & 0.047 & \dbar{0.000cm}{+17\%}{0.072cm} & 9.9*** & \gradcell{+0.41} & 0.032 & \dbar{0.088cm}{-21\%}{0.000cm} & -14.0*** & \gradcell{-0.57} & 0.041 & \dbar{0.000cm}{+2\%}{0.015cm} & 1.2 & \gradcell{+0.05} \\
Certainty & 0.015 & 0.013 & \dbar{0.057cm}{-14\%}{0.000cm} & -6.1*** & \gradcell{-0.25} & 0.010 & \dbar{0.153cm}{-36\%}{0.000cm} & -17.7*** & \gradcell{-0.78} & 0.009 & \dbar{0.174cm}{-41\%}{0.000cm} & -22.7*** & \gradcell{-0.98} \\
\rowcolor{gray!10}Causal reasoning & 0.019 & 0.024 & \dbar{0.000cm}{+22\%}{0.093cm} & 9.3*** & \gradcell{+0.36} & 0.033 & \dbar{0.000cm}{+71\%}{0.297cm} & 29.2*** & \gradcell{+1.17} & 0.028 & \dbar{0.000cm}{+46\%}{0.191cm} & 23.3*** & \gradcell{+0.94} \\
Insight & 0.023 & 0.035 & \dbar{0.000cm}{+50\%}{0.209cm} & 22.6*** & \gradcell{+0.70} & 0.034 & \dbar{0.000cm}{+47\%}{0.199cm} & 20.7*** & \gradcell{+0.67} & 0.023 & \dbar{0.015cm}{-3\%}{0.000cm} & -1.6 & \gradcell{-0.05} \\
\addlinespace[2pt]
\rowcolor{blue!10}\multicolumn{14}{@{}l}{\textit{\textbf{Affect and tone}}} \\[1pt]
Positive emotion & 0.060 & 0.062 & \dbar{0.000cm}{+3\%}{0.015cm} & 2.3* & \gradcell{+0.08} & 0.066 & \dbar{0.000cm}{+10\%}{0.040cm} & 7.7*** & \gradcell{+0.28} & 0.047 & \dbar{0.092cm}{-22\%}{0.000cm} & -20.8*** & \gradcell{-0.76} \\
\rowcolor{gray!10}Negative emotion & 0.020 & 0.029 & \dbar{0.000cm}{+46\%}{0.193cm} & 17.3*** & \gradcell{+0.47} & 0.023 & \dbar{0.000cm}{+16\%}{0.065cm} & 5.8*** & \gradcell{+0.17} & 0.021 & \dbar{0.000cm}{+3\%}{0.015cm} & 1.7 & \gradcell{+0.05} \\
Anxiety & 0.005 & 0.012 & \dbar{0.000cm}{+121\%}{0.420cm} & 18.3*** & \gradcell{+0.54} & 0.010 & \dbar{0.000cm}{+96\%}{0.403cm} & 13.8*** & \gradcell{+0.42} & 0.008 & \dbar{0.000cm}{+48\%}{0.204cm} & 12.1*** & \gradcell{+0.32} \\
\rowcolor{gray!10}Toxicity & 0.059 & 0.008 & \dbar{0.365cm}{-87\%}{0.000cm} & -11.5*** & \gradcell{-0.50} & 0.004 & \dbar{0.391cm}{-93\%}{0.000cm} & -12.2*** & \gradcell{-0.54} & 0.008 & \dbar{0.367cm}{-87\%}{0.000cm} & -11.6*** & \gradcell{-0.51} \\
Swearing & 0.001 & 5.7E$-$5 & \dbar{0.398cm}{-95\%}{0.000cm} & -17.3*** & \gradcell{-0.76} & 1.5E$-$5 & \dbar{0.414cm}{-99\%}{0.000cm} & -18.4*** & \gradcell{-0.82} & 1.1E$-$4 & \dbar{0.378cm}{-90\%}{0.000cm} & -16.2*** & \gradcell{-0.71} \\
\addlinespace[2pt]
\rowcolor{blue!10}\multicolumn{14}{@{}l}{\textit{\textbf{Social and instrumental}}} \\[1pt]
Social & 0.104 & 0.103 & \dbar{0.015cm}{-1\%}{0.000cm} & -0.8 & \gradcell{-0.02} & 0.094 & \dbar{0.040cm}{-9\%}{0.000cm} & -8.8*** & \gradcell{-0.20} & 0.082 & \dbar{0.089cm}{-21\%}{0.000cm} & -21.0*** & \gradcell{-0.51} \\
\rowcolor{gray!10}Affiliation & 0.015 & 0.027 & \dbar{0.000cm}{+80\%}{0.334cm} & 22.0*** & \gradcell{+0.62} & 0.022 & \dbar{0.000cm}{+42\%}{0.176cm} & 11.6*** & \gradcell{+0.33} & 0.018 & \dbar{0.000cm}{+16\%}{0.067cm} & 5.8*** & \gradcell{+0.17} \\
Drives & 0.075 & 0.108 & \dbar{0.000cm}{+44\%}{0.186cm} & 34.1*** & \gradcell{+1.15} & 0.089 & \dbar{0.000cm}{+19\%}{0.080cm} & 14.7*** & \gradcell{+0.51} & 0.081 & \dbar{0.000cm}{+8\%}{0.034cm} & 7.5*** & \gradcell{+0.26} \\
\rowcolor{gray!10}Achievement & 0.017 & 0.029 & \dbar{0.000cm}{+69\%}{0.288cm} & 24.5*** & \gradcell{+0.89} & 0.025 & \dbar{0.000cm}{+44\%}{0.186cm} & 16.5*** & \gradcell{+0.61} & 0.019 & \dbar{0.000cm}{+7\%}{0.028cm} & 3.3** & \gradcell{+0.12} \\
Work & 0.036 & 0.063 & \dbar{0.000cm}{+75\%}{0.315cm} & 30.9*** & \gradcell{+0.71} & 0.074 & \dbar{0.000cm}{+106\%}{0.420cm} & 39.2*** & \gradcell{+0.98} & 0.060 & \dbar{0.000cm}{+68\%}{0.286cm} & 28.6*** & \gradcell{+0.72} \\
\bottomrule
\end{tabular}}
\end{table*}

\subsection{Findings}
\label{sec:rq4-findings}

\subsubsection{Every system strips first-person voice, and Reddit Answers does so most strongly:}
Across all three systems, the largest and most consistent transformation is the loss of linguistic markers of personal testimony (Table~\ref{tab:rq4-register}). First-person singular pronouns fall from 3.3\% of tokens in the quoted community comments to 0.06\% in the Reddit Answers responses built from them ($d=-1.93$). Both general-purpose models move in the same direction on the same queries ($d=-1.77$ and $-1.32$). Personal pronouns more broadly fall from 9.5\% to 3.8\% in Reddit Answers ($d=-1.98$), while past-tense focus falls from 2.7\% to 0.9\% ($d=-1.43$). Toxicity and swearing also decline substantially across all three systems, and each system reduces expressions of certainty.

The fact that both general-purpose models show the same broad loss of first-person voice suggests that this transformation is not unique to Reddit Answers, but instead accompanies generative synthesis more generally. Reddit Answers nevertheless exhibits the strongest shift. Its use of first-person singular language falls below both \textit{GPT-4o-mini} and \textit{GPT-5} (between-system Cohen's $d=0.58$ and $1.22$, respectively). This produces a striking contrast: the system whose evidence base is explicitly composed of people's experiences retains the fewest linguistic cues that those claims originated as personal experiences. Together with RQ3, this also reveals a two-stage effect. Experiential voice is first less likely to survive selection, and much of what does survive is subsequently rewritten out of the first person.

Other dimensions do not move uniformly across models. Reddit Answers increases formality relative to its source comments (0.65 $\rightarrow$ 0.82, $d=+0.80$), and \textit{GPT-4o-mini} increases it further ($d=+1.31$), while \textit{GPT-5} lowers formality below the community baseline (0.65 $\rightarrow$ 0.54, $d=-0.58$). The two general-purpose models therefore differ substantially from one another despite belonging to the same model family. Positive emotion shows a similar divergence, with Reddit Answers and \textit{GPT-4o-mini} increasing it slightly while \textit{GPT-5} reduces it substantially ($d=-0.76$). These differences caution against treating every register change as an inherent consequence of generative synthesis. The consistent result across systems is specifically the weakening of first-person testimony.

\subsubsection{General-purpose generative search rarely draws on Reddit:}
The comparison with general-purpose LLMs also shows that community-grounded and general generative search draw on very different information environments. In the \textit{natural} condition, \textit{GPT-4o-mini} searched the web for only 3.6\% of queries and \textit{GPT-5} for 20.0\%, meaning that most advice-seeking queries were answered without external retrieval. Even when search was required, Reddit was almost absent from the resulting evidence. None of the 2{,}105 citations produced by \textit{GPT-4o-mini} under forced search pointed to Reddit, while only seven of the 2{,}065 citations from \textit{GPT-5} did. Across both models and search conditions, only 18 of 4{,}932 citations (0.4\%) pointed to Reddit, while institutional domains such as \texttt{irs.gov}, \texttt{cdc.gov}, \texttt{nhs.uk}, \texttt{mayoclinic.org}, and \texttt{pmc.ncbi.nlm.nih.gov} were much more common.

The experiential and community-authored knowledge that constitutes Reddit Answers' evidence base is therefore largely absent from the information sources used by general-purpose generative search in our setting. The systems are not simply different interfaces over the same evidence. Reddit Answers provides access to a body of community discourse that general-purpose systems rarely retrieve, making the routing and selection decisions identified in RQ2 and RQ3 particularly consequential for what users ultimately see.

\begin{takeawaybox}
\textbf{\faLightbulbO ~\large Takeaway:}
Generative synthesis substantially changes how community evidence is presented. Across systems, first-person testimony is consistently weakened, with Reddit Answers showing the strongest shift despite being grounded directly in community-authored experiences. At the same time, general-purpose generative search rarely retrieves Reddit at all.
\end{takeawaybox}

\section{Discussion}

We now discuss what our findings imply for the design of systems that retrieve over community or group discourse, for the growing class of research systems built on the same premise, and for how audits of deployed generative search should be conducted.

\subsection{Generative Search Adds Another Layer of Ranking}

One of the key findings of our audit is that retrieval largely determines the final Reddit Answers response. When repeated queries retrieve the same sources, Reddit Answers produces almost the same answer, while when retrieval changes, the answer changes with it. This places a lot of weight on steps 2 and 3 in \autoref{fig:reddit-answers-pipeline}, since the generator can only summarize the material that the retrieval and selection stages have already allowed through.

We find that these stages are not unbiased. Within a thread, score is the strongest predictor of whether a comment is selected, and top-level comments and earlier posted comments are also substantially more likely to appear. The comments that make it into an answer are therefore systematically more visible than those left out. This matters because visibility on social platforms is itself the outcome of earlier ranking, position, and social influence processes. Prior work has shown that early feedback and ranking can shape what later users see and reward~\cite{muchnik2013social,lerman2014leveraging,chan2025ranking}. Reddit Answers does not replace or try to correct those dynamics, instead placing another selection layer on top of them.

This distinction is important because much of the current push in AI focuses on making models more capable, including the models that produce generative search answers. Yet a better generator cannot recover a perspective that was never retrieved. Improvements to prompting, attribution, or answer formatting can change how selected evidence is presented, but they cannot change whose evidence had the opportunity to appear in the first place. Our results therefore suggest that improving community-grounded search requires attention further upstream. Work on fairness in retrieval-augmented generation makes a similar point by treating exposure at retrieval time as a central part of representational fairness~\cite{wu-etal-2025-rag,tikoo2026towards}.

For Reddit, this creates a concrete design opportunity. Rather than relying only on a generic retriever optimized for relevance, the platform could train or fine-tune a retrieval model specifically for community discourse, similar to how research on community-governance and modular pluralism has tackled transferability challenges with general-purpose models~\cite{park-etal-2021-detecting-community,feng-etal-2024-modular,zhan-etal-2025-slm,goyal-etal-2025-momoe}. Its objective could reward not only semantic fit, but also coverage of substantively different perspectives, representation across communities, and exposure to contributions that are informative even when they are not already highly visible. Such an objective would not require treating every comment equally. It would instead give the system a way to separate what is relevant from what is merely already visible, and to avoid treating popularity as a proxy for consensus. Our findings suggest that this is the stage where interventions to preserve plurality are likely to matter most.

\subsection{Cross-Community Retrieval Is an Editorial Decision}

When a user asks a question through Reddit Answers, they may want useful perspectives from across the platform, and the retrieval process must decide which communities get to contribute to the answer. We find that a typical answer draws from roughly seven communities. The system therefore creates what we might call a \textit{``synthetic community context''} for every query. This context did not exist before the search, but was rather assembled by the retriever from communities with different norms, expertise, and expectations about what constitutes useful advice~\cite{chandrasekharan2018internet,weld_what_2022}. Cross-community retrieval is therefore not only a decision about the relevance, but also an \textit{editorial decision} about which parts of Reddit get to answer a particular question.

Our results show that this assembly process is systematic. Retrieval tends to route between communities of similar size, while questions originating in smaller communities are often answered using material from larger ones. We do not argue that this movement is necessarily undesirable. Drawing from several communities may expose users to useful perspectives that they would never have encountered by browsing a single subreddit. However, these choices currently happen invisibly. Our origin posts give us a useful probe here rather than a normative ground truth. They provide one source that we know is relevant to each query. Yet for 79.3\% of queries, none of the retrieved posts is a closer semantic match than this known relevant source. The communities assembled around a query therefore cannot be understood simply as the places where the system found better matching content.

This creates an interesting contrast with recent HCI work on socially grounded generation such as Social-RAG~\cite{10.1145/3706598.3713749}, which treats the history and social signals of a particular group as important context that an AI system should preserve when generating for that group. Systems such as Reddit Answers face almost the opposite design problem. It draws from several communities and turns their contributions into a single response. The challenge is therefore not only how to ground generation in social context, but how to combine several social contexts without making their differences disappear. HCI has long explored interfaces that help users navigate multiple viewpoints, from NewsCube's organization of competing perspectives~\cite{10.1145/1518701.1518772} to work showing that how disagreement is presented can shape engagement with diverse information~\cite{10.1145/1753326.1753543}. Similarly, more recent systems such as CommSense treat the presentation of online comments as a design problem rather than a neutral display choice~\cite{10.1145/3772318.3790530}. Community-grounded search could build on this tradition by making the communities behind an answer more visible, letting users steer whether they want broad or community-specific perspectives, or explicitly retrieving complementary and minority perspectives rather than letting the most visible communities dominate the synthetic context. Researchers could then test these systems through a user study or via empirical audits like ours. The goal of such a system could be to make the construction of its synthetic community context more deliberate and legible to users.

This also opens a direct avenue for future HCI research. A user study could compare how people make the same advice-seeking decision when they read the underlying Reddit threads, receive a synthesized cross-community answer, or receive an answer from a general-purpose LLM. Such a study could examine not only which decision people make, but also their confidence, reliance on the system, perception of community consensus, and understanding of where the advice came from. Prior work already shows that AI summaries of social media discussions can shift perceived majority opinion and perceived balance~\cite{10.1145/3772318.3790945}. More broadly, human-AI decision-making research has shown that seemingly small choices in how AI mediation is presented can change reliance and decision behavior~\cite{10.1145/3411764.3445717,10.1145/3593013.3594087}. Community-grounded generative search gives us a new version of this question: \textit{How does making a decision from a synthetic representation of several communities differ from encountering those communities themselves?}

\subsection{Synthesis Turns Testimony into Advice}

For advice and support communities, first-person language is not merely a writing style; rather it tells the reader what kind of evidence they are encountering. For example, ``This happened to me'' carries a different epistemic meaning from ``this is what you could do.'' Prior HCI research has shown that storytelling, self-disclosure, and personal experience play important roles in how people provide advice, empathy, and support online~\cite{10.1145/3290605.3300574,andalibi2016understanding,andalibi2017sensitive,sharma2018mental}. When a system rewrites these accounts, it can therefore change more than their wording.

Our findings show that this happens twice. Comments with stronger experiential voice are less likely to be selected in the first place. The experiences that do survive selection are then largely rewritten out of the first person. First-person singular language falls from 3.3\% of tokens in the source comments to 0.06\% in Reddit Answers, a larger shift than we observe for either general-purpose LLM we tested. The system therefore does not simply summarize lived experience, but changes lived experience into the voice of generalized advice.

We view this as a form of \textit{evidentiary flattening}. A statement grounded in one person's experience can emerge from synthesis in the same narrator voice as a statement supported across many comments. The underlying claim may remain similar, but an important cue about where that claim came from has been weakened. This creates an interesting contrast with recent work on synthetic lived experience, which shows that LLMs can adopt first-person language that falsely implies experiences they never had~\cite{goel2026ai}. Here we observe the inverse problem. AI mediation can take authentic lived experience and make it sound less like lived experience. Both cases blur the relationship between a claim and the person or evidence behind it. This distinction also matters because summaries can shape how readers understand the discussion being summarized or the decisions they make. \citet{10.1145/3772318.3790945} show that AI-generated summaries of social media conversations can change perceptions of majority opinion and of how balanced a discussion is. 

For community-grounded search, provenance should therefore be treated as a design objective rather than a citation feature. HCI researchers and system designers could build interfaces that preserve the evidentiary role of source material, for example by keeping first-person accounts visibly attributed, separating individual experiences from patterns repeated across sources, or showing when a recommendation reflects broad agreement rather than a small number of voices. Such designs may also help preserve a role for users' own judgment. \citet{shaw2026thinking}'s Tri-System Theory describes generative AI as a form of external ``System 3'' cognition and warns that users may engage in cognitive surrender when they adopt its outputs without sufficient deliberation. Community-grounded search offers an opportunity to design against this tendency. Rather than presenting synthesis as the endpoint, interfaces could invite users back into the underlying evidence through salient quotations, contrasting perspectives, or lightweight paths to inspect the original discussion. Such interventions could help move users from simply accepting a synthesized answer toward more deliberate engagement with its sources. The design challenge is therefore not just to show where a claim came from, but to preserve what kind of evidence it was and give users meaningful opportunities to engage with it.

\subsection{Generative Search Audits Must Account for Run-to-Run Variation}

A generative search response is not a fixed system output. It is one draw from a distribution. We find that identical queries often retrieve substantially different sources, even when the resulting answers remain linguistically similar. This instability is also corpus-dependent. Smaller communities are considerably less reproducible than larger ones. Recent work on Google AI Overviews and Gemini reports a similar lack of consistency across repeated runs of the same query~\cite{10.1145/3805712.3809667}, while emerging work on generative-search measurement similarly argues that single-run visibility estimates can give a misleadingly precise picture of system behavior~\cite{sielinski2026quantifying}.

This changes how these systems should be audited. Future audits should repeat queries, quantify source overlap across runs, and propagate this variation into reported uncertainty. They should also sample across different parts of the corpus. An audit restricted to large or popular communities could systematically overstate the stability of the same underlying system. The relevant unit of analysis is therefore not a single answer, but the distribution of evidence the system repeatedly chooses to surface.

Furthermore, this variability may also become something designers can use. A system could estimate retrieval stability before synthesis and surface a lightweight signal to the user when the evidence behind an answer changes substantially across repeated runs. Whether users would interpret such a signal as uncertainty, disagreement, or unreliability remains a question for future research, since prior HCI work shows that uncertainty representations do not necessarily convey the meaning designers intend~\cite{10.1145/2858036.2858558,8457476}. More broadly, our findings suggest that community-grounded search should make not only its sources visible, but also the uncertainty in how those sources were selected.

\subsection{Limitations and Future Work}

Our work has limitations, which point towards opportunities for exciting future work.

\subsubsection{Observational versus Causal Insights:} Our study is observational and should not be interpreted causally. This is consistent with a long tradition of black-box algorithm audits, which characterize systematic behavior by repeatedly querying deployed systems and observing their outputs when their internal models and ranking mechanisms are inaccessible \cite{metaxa2021auditing,10.1145/3449148,sandvig2014auditing}. Establishing causal effects in our setting would require independently manipulating properties such as comment visibility, position, or linguistic register while holding relevance, content, and the surrounding discussion fixed. These properties arise jointly through user behavior on a live platform and cannot be cleanly intervened on through the Reddit Answers interface. Our analyses therefore identify systematic associations in what the system retrieves and selects, rather than estimating the effect that changing any single feature would have on selection.

\subsubsection{Longitudinal Analysis:} Our repeated-query design is intended to measure short-term system stability rather than longitudinal change. We submit identical queries in closely spaced runs so that differences are more likely to reflect the system's own non-determinism rather than changes in Reddit itself. This means we do not study how answers evolve over longer periods as new posts and comments appear, votes accumulate, communities change, or the retrieval index and underlying models are updated. A longitudinal audit could reveal additional forms of instability and representational drift that our design does not capture.

\subsubsection{Empirical Audit versus User Studies:} Our analysis focuses on what the system retrieves and presents, not on how users ultimately interpret or act on those outputs. We can show that first-person testimony is reduced, that community context is recombined, and that some perspectives receive less exposure, but we do not measure whether these changes alter trust, understanding, perceived consensus, or decision quality. These remain important user-facing questions for future HCI work. Our study also examines one deployed platform and a set of advice- and support-seeking communities, so the specific patterns we observe may not generalize to other generative search systems, platforms, or domains.

\subsubsection{Demographically Neutral Queries:}
\label{sec:limitations-demographics}

Our query-conversion procedure generally removes explicit demographic markers such as age, gender, and nationality. We make this choice to hold demographic framing relatively constant while auditing how Reddit Answers routes, selects, and synthesizes evidence. This necessarily abstracts away part of how people actually seek advice on Reddit, where questions often contain personal and social context~\cite{chen202430f}. Prior work shows that demographic attributes can produce disparities in both the retrieval and generation stages of RAG systems~\cite{wu-etal-2025-rag}, and that changing the role or social context implied by otherwise similar questions can systematically alter LLM responses~\cite{kaur-etal-2026-whos}. Our findings therefore characterize system behavior for demographically neutralized information needs, rather than the full range of situated queries users may pose. A natural extension is a controlled demographic-perturbation audit in which the same underlying query is varied along demographic or role attributes to measure how those cues change community routing, source selection, and synthesis.

\section{Ethical Considerations}
\label{sec:ethics}

Our study analyzes publicly accessible Reddit content and the outputs of Reddit's deployed AI search system. We do not interact with Reddit users, intervene in community discussions, or post experimental content. For user-authored posts, comments, scores, and metadata, we use Reddit's official API rather than scraping rendered Reddit pages, and adhere to the access and rate limits imposed by the API. We use these data only for the analyses described in this work and report results primarily in aggregate to minimize unnecessary exposure of individual users.

Collecting Reddit Answers requires a different access mechanism because Reddit does not provide a research API for AI-search responses. We submit queries to the publicly available Reddit Answers interface and record the system-generated response and its source references. Reddit explicitly permits users to share AI-search responses externally\footnote{\href{https://support.reddithelp.com/hc/en-us/articles/32026729424916-Reddit-s-AI-search\#h\_01JCE0HZKGE0DY90FRFJ3A7YKH}{Reddit's AI search FAQs}}, which allows us to reproduce and analyze the generated outputs reported in this paper. Our querying was rate-limited to avoid imposing undue load on the service, and we did not attempt to circumvent authentication, access controls, or any other technical protections.

\section{Conclusion}

Generative search over online communities does more than summarize what people have said. It decides which communities are represented, which contributions are surfaced, and how those contributions are rewritten into a single answer. Our audit shows that these choices systematically favor already-visible content, reshape the social context around a query, and weaken cues that distinguish lived experience from generalized advice. We argue that community-grounded search should therefore be designed not only for relevance and fluency, but also for provenance, plurality, and legibility. More broadly, our findings highlight the need to treat retrieval and synthesis as sociotechnical choices that shape how collective knowledge is represented.


\bibliographystyle{ACM-Reference-Format}
\bibliography{references}

\appendix
\section{Robustness of RQ3 Selection Analyses}
\label{app:rq3-robustness}

Our RQ3 analyses are observational. We observe which comments Reddit Answers surfaces and which comments in the same retrieved threads it passes over, but we do not observe the system's internal ranking criteria. One plausible unmeasured factor is particularly important to address. A comment may be selected because it directly answers the user's question, and this property may itself correlate with many of the features we study. Direct answers may be longer, more likely to contain links, more visible, and written in a more formal or prescriptive register. If so, some associations attributed to these observable features could instead reflect unmeasured differences in how directly comments answer the query.

We address this concern in two complementary ways. First, we remove one major structural source of variation by restricting the analysis to top-level comments, which all respond directly to the source post rather than to another commenter. Second, we use Oster's coefficient-stability method to ask how strong selection on an unmeasured confounder would need to be to explain away the remaining associations. Neither analysis identifies a causal effect, but together they help distinguish the findings that remain stable under this alternative explanation from those that do not.

\subsection{Restricting Comparisons to Top-Level Comments}

\subsubsection{Design:}
Reply depth provides the clearest structural route through which direct answer relevance could confound our estimates. Top-level comments respond directly to the source post, while nested comments frequently respond to another comment and may instead express agreement, disagreement, clarification, or support. We therefore refit the RQ3 models using only top-level comments. This leaves approximately 220{,}600--241{,}100 comments across 19{,}300--20{,}300 threads per draw, covering roughly 90\% of the threads used in the primary analysis. We recompute score percentile within this restricted sample and otherwise retain the original model specifications.

\subsubsection{Structural results remain stable:}
The strongest structural associations are nearly unchanged. Score percentile remains the dominant predictor of selection (OR 2.86, compared with 2.88 in the primary analysis), while comment length (OR 1.77 versus 1.79) and age (OR 0.72 versus 0.73) are similarly stable. Moderator-distinguished comments (OR 0.013) and stickied comments (OR 0.052) also remain rarely selected.

The association with external links weakens from OR 2.25 to 1.591. This suggests that part of the link association is related to comment position, although a substantial positive association remains after restricting comparisons to comments that all respond directly to the post.

\subsubsection{The main language results also persist:}
The two language findings most central to our interpretation remain robust. Formality attenuates but remains the strongest linguistic predictor of selection (OR 1.334, [1.289, 1.381], $q<10^{-4}$), while experiential voice remains negatively associated with selection (OR 0.869, [0.841, 0.898], $q<10^{-4}$). Prosocial warmth also remains negatively associated with selection (OR 0.867), while directive prescription remains positively associated (OR 1.059).

Two additional associations become stronger under the restricted specification. Toxicity changes from no overall association in the primary model to a modest negative association (OR 0.963, $q=0.025$; visibility-adjusted OR 0.951, $q=0.003$), while the penalty associated with Reddit's \texttt{controversial} flag strengthens from OR 0.744 to 0.582. These results are consistent with the broader finding that contested or undesirable content is less likely to be surfaced.

\subsubsection{Several smaller effects do not replicate:}
Not every result survives this restriction. The associations for \textit{certainty} and \textit{hedging}, which were statistically significant in the primary model, shrink to OR 1.005 [0.985, 1.025] and OR 1.002 [0.984, 1.021], respectively. Informational specificity behaves similarly. We therefore do not interpret these features as stable selection preferences in the main text.

Positive affect remains directionally consistent but attenuates substantially, from OR 0.801 to 0.961. We consequently treat this result more cautiously than the stronger findings for experiential voice and formality. The apparent penalty on the OP's own comments also does not replicate among top-level comments (OR 0.842, $q=0.61$), suggesting that the primary association largely reflects where OPs tend to participate in the thread rather than a stable preference against the author of the source post.

\subsection{Sensitivity to Unmeasured Confounding}

\subsubsection{Bounding the remaining concern:}
Restricting the analysis to top-level comments removes one important structural difference, but it does not directly observe whether a comment substantively answers the query. We therefore complement the restricted analysis with Oster's coefficient-stability approach~\cite{oster2019unobservable,altonji2005selection}.

Oster's method compares how much an estimated coefficient changes as observed controls are added against how much additional variance those controls explain. The resulting $\delta$ measures how strong selection on unobserved factors would need to be relative to selection on observed factors to reduce the coefficient to zero. Values with $|\delta|>1$ indicate that unobserved selection would need to be stronger than observed selection under the proportional-selection assumption.

Because the method requires an $R^2$ that is not directly available from conditional logistic regression, we refit the relevant specifications as linear probability models after absorbing thread fixed effects through within-thread demeaning. We only interpret the sensitivity result when the linear model reproduces the direction of the corresponding conditional-logit estimate. We set $R^2_{\max}=1.3R^2_{\text{long}}$ and compute $\delta$ separately for each draw.

\subsubsection{The central associations require substantial unmeasured selection to disappear:}
Adding the visibility controls increases the within-$R^2$ from 0.025 to 0.130, providing enough movement for the coefficient-stability calculation to be informative. Experiential voice yields $\delta=6.0$, ranging from 5.5 to 6.5 across draws. Under the proportional-selection assumption, selection on unobserved factors would therefore need to be approximately six times as strong as selection on the observed controls to reduce this association to zero.

Directive prescription yields $\delta=5.9$, while formality yields $\delta=2.7$. Among the structural features, links yield $\delta=13.7$ and comment length $\delta=3.4$. Prosocial warmth and toxicity produce negative values ($-8.4$ and $-7.6$), reflecting that their coefficients move away from zero rather than toward it when the observed controls are introduced.

These results do not rule out unmeasured confounding. They instead quantify how strong unobserved selection would need to be, relative to the observed visibility-related factors, to eliminate the associations on which our main interpretation relies.

\subsubsection{The same weaker effects remain fragile:}
The Oster analysis also reinforces our decision not to interpret several smaller associations. Certainty, hedging, and informational specificity all produce $\delta$ values below one, with ranges that include zero. These are also the features that disappear in the top-level-only analysis. We therefore treat them as insufficiently robust and do not use them to support conclusions about Reddit Answers' preferred epistemic stance.

Positive affect is less clear. Although its association survives the top-level restriction, it attenuates substantially, and the linear probability specification used for the Oster analysis does not reproduce the conditional-logit direction. We therefore do not report an Oster bound for this feature and avoid placing substantial interpretive weight on it.

\subsection{What These Checks Establish}

These robustness analyses do not remove the possibility of unmeasured confounding, nor do they turn the observational audit into a causal design. Directly identifying the effect of a comment's language or structure would require an intervention that changes one property while holding its relevance and surrounding discussion fixed. Such an intervention is not available through the deployed Reddit Answers interface and would likely require manipulating content posted to the communities themselves.

The checks instead help distinguish robust descriptive patterns from fragile ones. The associations on which our main interpretation rests remain stable when comparisons are restricted to comments that all reply directly to the same post, and the Oster bounds indicate that relatively strong unobserved selection would be required to eliminate them. In particular, the strong role of community visibility, the preference for formal register, and the penalty associated with experiential voice survive both analyses. Smaller effects including certainty, hedging, informational specificity, and OP status do not, and we therefore avoid drawing conclusions from them in the main paper.

\section{Source Support of Generated Claims}
\label{app:rq4-support}

Reddit Answers presents its responses as grounded in community evidence. Inline quotations are linked to specific comments, while the surrounding prose summarizes, combines, and extends those sources into a coherent answer. We therefore ask a more granular question than whether the response is generally related to its sources: whether the individual claims made in the generated answer can actually be supported by the comments Reddit Answers attributes to it.

\subsection{Claim-Level Evaluation}

\subsubsection{Decomposing answers into atomic claims:}
Generated answers often pack several distinct propositions into a single sentence. For example, a sentence such as ``You should call your provider because the charge may be reversible and other users report success after disputing it'' contains at least three separable claims: that the user should contact the provider, that the charge may be reversible, and that other users report successful disputes. Evaluating the sentence as a single unit would make it unclear which part is supported and which is not.

We therefore decompose each Reddit Answer into \textit{atomic claims}, defined as the smallest self-contained factual, descriptive, or advisory propositions that can be evaluated independently against evidence. A claim may describe what users report, characterize a pattern across comments, make a factual statement, or offer a recommendation. We exclude purely structural text such as headings and transitions that do not make a substantive claim.

On a stratified subset of 2{,}000 atomic claims, with 1{,}000 from each community set and balanced across the four domains, we evaluate whether each claim is supported by the evidence available to Reddit Answers in a single draw.

\subsubsection{Constructing evidence at three attribution levels:}
For every claim, we construct three evidence sets that mirror how Reddit Answers exposes or uses its sources. The first contains the comments quoted inline in the answer. The second contains comments that Reddit Answers lists as sources but does not quote directly. The third contains the eight comments from the retrieved threads that are most semantically similar to the claim but were not surfaced in the answer. This gives us progressively broader evidence. The first tier corresponds most closely to what a user can directly inspect from the answer itself. The second asks whether support exists elsewhere among the sources the system explicitly attributes to the response. The third asks whether support exists in the retrieved context even when the system does not surface it.

\subsubsection{Using two independent verifiers:}
We evaluate each atomic claim against each evidence tier using two separate verification methods. The first is \textit{Bespoke-MiniCheck-7B}~\cite{tang-etal-2024-minicheck}, a model designed for factual consistency evaluation. The second is a \textit{GPT-4o-mini}~\cite{hellogpt4o} (\texttt{gpt-4o-mini-2024-07-18}) LLM judge prompted independently to determine whether the evidence supports the claim. We use two verifiers because source-support evaluation is itself noisy. MiniCheck reports 77.4\% balanced accuracy on the LLM-AggreFact benchmark~\cite{tang-etal-2024-minicheck}, which is useful but not sufficient for treating its output as ground truth. Prior work also shows that factuality metrics can disagree substantially, especially when claims are heavily paraphrased or supported by information spread across multiple passages~\cite{godbole-jia-2025-verify}. Both conditions are common in Reddit Answers, which synthesizes and rewrites multiple comments rather than reproducing them verbatim.

\subsection{Findings}

\subsubsection{Verifier disagreement is substantial:}
The two verification methods produce noticeably different unsupported-claim rates. MiniCheck flags 28.8\% of claims as unsupported by any evidence tier, compared with 11.0\% for the LLM judge. Agreement is moderate when evaluating the comments quoted inline ($\kappa=0.42$), but falls to nearly zero for retrieved but unsurfaced evidence ($\kappa=0.09$).

Given this disagreement, we do not treat either verifier as authoritative; instead, we focus on the claims for which the two methods agree.

\subsubsection{A non-trivial subset of claims is unsupported under both verifiers:}
Across the 2{,}000 claims, 8.7\% are classified as unsupported by both verification methods. At the other extreme, 31.1\% are classified as unsupported by at least one verifier. The true rate cannot be identified from these measurements alone, but the overlap provides a conservative signal that some generated claims cannot be readily traced to the evidence Reddit Answers attributes to them.

The ordering across domains is more stable than the absolute rates. Finance produces the highest unsupported rate under both verifiers, at 37.5\% under MiniCheck and 13.5\% under the LLM judge. This consistency suggests that domain differences may be more reliable than the exact system-wide prevalence.

\section{LLM Prompts}
\label{app:prompts}

\begin{tcolorbox}[
    breakable,
    colback=black!2,
    colframe=black!35,
    title={S3 --- relevance gate (\textit{gpt-5.4-nano-2026-03-17})},
    fonttitle=\bfseries
]
\small\ttfamily\raggedright
You are screening a Reddit self-post to decide whether it represents a genuine
advice-, support-, or information-seeking request that a real person could
plausibly type into a question-answering search tool.\\[5pt]
Decide KEEP or DROP.\\[5pt]
KEEP if the post's core is a person seeking guidance, information, opinions,
others' experiences, reassurance, or help making a decision, AND that need
could be expressed as a standalone question understandable without the post's
specific backstory.\\[5pt]
DROP if the post is primarily any of:\\
- news, article, or link sharing\\
- a success story, update, or progress report with no open question\\
- venting or journaling with no answerable request\\
- off-topic, meme, or joke\\
- meta, subreddit administration, or moderation\\
- self-promotion, survey, recruitment, or solicitation\\
- a poll\\
- so idiosyncratic or context-bound that no generalizable question remains\\[5pt]
Output JSON only, no prose:\\
\{''keep'': \textless{}true|false\textgreater{}, ''reason'': ''\textless{}$\leq$15 words\textgreater{}''\}
\end{tcolorbox}

\begin{tcolorbox}[
    breakable,
    colback=black!2,
    colframe=black!35,
    title={S4 --- query conversion (\textit{gpt-5.4-nano-2026-03-17})},
    fonttitle=\bfseries
]
\small\ttfamily\raggedright
You rewrite a Reddit self-post into a single, natural, standalone question---the kind a real person would type into a search or AI-answers box---preserving
the poster's actual need while removing identifying and hyper-specific detail.\\[5pt]
Rules:\\
1. Output ONE first-person question (or a 1--2 sentence request) that captures the post's core need.\\
2. Add nothing. Do not introduce specifics, entities, numbers, or constraints not present in the original. Never invent detail.\\
3. Remove personally identifying information (names, usernames, employers, precise locations, unique dates). Strip demographic specifics (exact age, gender, nationality) UNLESS they are the crux of the question. Base queries should be demographic-neutral.\\
4. Abstract one-off backstory so the question generalizes, but keep context that is essential to the need (keep ''after a layoff'' if central; drop ''at my cousin's wedding in Ohio'').\\
5. Natural register---how a person actually types, not formalized prose. Concise: $\leq$ \textasciitilde{}40 words.\\[5pt]
Output JSON only, no prose:\\
\{''normalized\_query'': ''\textless{}string|null\textgreater{}'',\\
 ''note (optional)'': ''\textless{}$\leq$15 words\textgreater{}''\}
\end{tcolorbox}

\subsection{Validation of the S3 Relevance Gate}
\label{app:s3-validation}

S3 is intended as a high-precision filtering step, since its purpose is to ensure that posts carried forward into query construction genuinely express an advice-, support-, or information-seeking need. We therefore constructed a 100-post validation sample stratified by source community, with five posts sampled from each of the 20 communities, and balanced the sample across the model's decisions (50 \textsc{Keep} and 50 \textsc{Drop}). The posts were shuffled, and two annotators independently applied the same binary inclusion criterion used by the LLM while blind to its decisions.

Our primary validation focuses on the 50 posts passed by S3, since these are the posts that enter downstream query construction. The two annotators independently judged 84\% and 88\% of these posts, respectively, to satisfy the inclusion criterion. They agreed on 44 of the 50 cases (88\% raw agreement; Cohen's $\kappa\approx0.50$ indicating moderate agreement), with 40 posts accepted by both annotators. We prioritize this precision-oriented check because false positives would introduce out-of-scope queries into the audit, whereas posts rejected by S3 do not enter any further analyses.

\section{Manual Fidelity Audit of Query Conversion}
\label{app:query-validation}

Because the S4 conversion from a narrative Reddit post to a standalone query has no unique gold-standard output, we evaluate it through a manual fidelity audit. The first author randomly sampled 60 source-post/query pairs, stratified by community, from the final dataset and inspected whether each generated query (1) preserved the post's central information-seeking need, (2) introduced any unsupported substantive information, and (3) retained identifying or unnecessary demographic detail that should have been removed. Across the 60 inspected pairs, we found no cases in which the generated query materially changed the underlying information need or introduced unsupported substantive content. We consider this audit to be a qualitative sanity check on the query extraction, rather than as an estimate of the full-corpus error rate.

\section{Query Conversion Examples}
\label{app:examples}

Table~\ref{tab:examples} shows two distilled queries from each of five communities per set, chosen to span the four domains. Only the queries are shown: they are what we submit to the system, and unlike the posts they come from they are abstracted and demographically neutral by construction.

\begin{table*}[ht]
    \centering\small
    \sffamily
    \caption{Examples of two distilled queries from each of ten source communities (five from the \textsc{Large} set and five from the \textsc{Small} set)}\Description{Table showing examples of distilled queries used for the audit.}
    \label{tab:examples}
    \setlength{\tabcolsep}{5pt}
    \renewcommand{\arraystretch}{1.15}
    \begin{tabular}{@{}p{2.2cm} p{5cm} p{2.5cm} p{5cm}@{}}
    \multicolumn{2}{c}{\textbf{\textsc{Large} communities}} & \multicolumn{2}{c}{\textbf{\textsc{Small} communities}} \\
    \cmidrule(lr){1-2}\cmidrule(lr){3-4}
    \rowcolor{blue!10}\textbf{Community} & \textbf{Distilled query} & \textbf{Community} & \textbf{Distilled query} \\
    \midrule
    \multirow{2}{=}{\textbf{r/personalfinance}} & How should we invest our money to grow savings more efficiently? & \multirow{2}{=}{\textbf{r/PersonalFinanceNZ}} & What are the best options for kids savings accounts? \\*
     & What's the best high-yield savings account for growing my money? &  & How long are you fixing your mortgage rate right now? \\
    \addlinespace[3pt]
    \rowcolor{gray!10}\multirow{2}{=}{\textbf{r/AskDocs}} & Could my symptoms be a seizure, and what should I do next? & \multirow{2}{=}{\textbf{r/AusLegal}} & Where can I get a contractor service agreement drafted cheaply? \\*
    \rowcolor{gray!10} & How concerning is a WBC of 28.6 in someone who's been sweating at night? &  & Where can I learn about my country's law if I don't want to go to uni? \\
    \addlinespace[3pt]
    \multirow{2}{=}{\textbf{r/techsupport}} & What is the D: drive on my computer? & \multirow{2}{=}{\textbf{r/AskTechnology}} & Do Oppo Enco Buds 2 sound better than Buds 3 Pro? \\*
     & Can I run PC disc games using an external CD/DVD drive over USB? &  & What is the point of Reddit negative comment bots? \\
    \addlinespace[3pt]
    \rowcolor{gray!10}\multirow{2}{=}{\textbf{r/careerguidance}} & How common is it to earn over \$100k a year in the US? & \multirow{2}{=}{\textbf{r/UKJobs}} & What would make your job easier, big or small? \\*
    \rowcolor{gray!10} & Can I become a programmer if I'm already in my late 30s? &  & Is it ever worth returning to a previous employer? \\
    \addlinespace[3pt]
    \multirow{2}{=}{\textbf{r/HealthAnxiety}} & What's helped the most with your health anxiety? & \multirow{2}{=}{\textbf{r/MentalHealthUK}} & How do people cope with body dysmorphia (BDD)? \\*
     & How long was your longest health anxiety flare-up? &  & Do you have ``BPD rage,'' and if so how do you cope and manage it? \\
    \bottomrule
    \end{tabular}
    \end{table*}

\end{document}